\documentclass[fleqn,usenatbib]{mnras}

\usepackage{newtxtext,newtxmath}
\usepackage{caption}

\usepackage[T1]{fontenc}
\usepackage{xcolor}

\DeclareRobustCommand{\VAN}[3]{#2}
\let\VANthebibliography\thebibliography
\def\thebibliography{\DeclareRobustCommand{\VAN}[3]{##3}\VANthebibliography}

\usepackage{graphicx}	
\usepackage{amsmath}	

\newcommand{\pypeit}{\texttt{PypeIt}}
\newcommand{\band}{$B^{2}\Sigma^{+}-X^{2}\Sigma^{+}$}
\newcommand{\axband}{$A^{2}\Pi_{\rm u}-X^{2}\Sigma^{+}$}
\newcommand{\lsfg}{$LSF(\lambda)_{\rm G}$}
\newcommand{\lsfv}{$LSF(\lambda,\gamma=0)_{\rm V}$}
\newcommand{\ronezero}{N(1)/N(0)}
\newcommand{\rtwoone}{N(2)/N(1)}

\title[The CMB temperature]{A determination of the cosmic microwave background temperature using Galactic molecules}

\author[Cooke \& Welsh]{
Ryan J. Cooke,$^{1,2}$\thanks{E-mail: ryan.j.cooke@durham.ac.uk (RJC)}
Louise Welsh$^{3,4,5}$
\\
$^{1}$Centre for Extragalactic Astronomy, Department of Physics, Durham University, South Road, Durham DH1 3LE, UK\\
$^{2}$Department of Physics, Durham University, South Road, Durham DH1 3LE, UK\\
$^{3}$INAF - Osservatorio Astronomico di Trieste, via G. B. Tiepolo 11, I-34143 Trieste, Italy\\
$^{4}$Dipartimento di Fisica G. Occhialini, Universit\`a degli Studi di Milano Bicocca, Piazza della Scienza 3, I-20126 Milano, Italy\\
$^{5}$IFPU - Institute for Fundamental Physics of the Universe, via Beirut 2, I-34151 Trieste, Italy\\
}

\date{Accepted XXX. Received YYY; in original form ZZZ}

\pubyear{2024}

\begin{document}
\label{firstpage}
\pagerange{\pageref{firstpage}--\pageref{lastpage}}
\maketitle

\begin{abstract}
We report a new, reliable determination of the CN excitation temperature of diffuse molecular clouds in the Milky Way, based on ultra high spectral resolution observations. Our determination is based on CN \band\ (0,0) vibronic band absorption spectra seen along the lines of sight to eight bright Galactic stars. Our analysis is conducted blind, and we account for multiple sources of systematic uncertainty. Like previous studies, our excitation temperature measures exhibit an intrinsic scatter that exceeds the quoted uncertainties. Accounting for this scatter, we derive a $3\%$ determination of the typical CN excitation temperature, $T_{01}=2.769^{+0.084}_{-0.072}~{\rm K}$, which is consistent with the direct determination of the cosmic microwave background temperature. We also perform a single joint fit to all sightlines simultaneously, and find that our data can be fit with an excitation temperature $T_{01}=2.725\pm0.015~{\rm K}$ --- a $0.55\%$ measure that is consistent with the CMB temperature. We propose a future observational strategy to reduce systematic uncertainties and firmly test the limitations of using CN as a cosmic microwave background thermometer.
\end{abstract}



\section{Introduction}

The serendipitous discovery of the cosmic microwave background (CMB) by \citet{PenziasWilson1965}, and the subsequent demonstration that this background is almost a pure blackbody \citep{Mather1990,Gush1990}, were two important discoveries that set the stage for the Hot Big Bang cosmological model. Since these early works, the CMB has been observed by multiple experiments that have helped to refine the details of our cosmological model. For the most part, these experiments have focused on measuring the angular power spectrum of fluctuations that reflect the density inhomogeneities during photon decoupling. The temperature fluctuations that we record today are measured relative to the present day CMB temperature, which is one of the best-measured cosmological quantities. The currently accepted value of the CMB temperature, $T_{\rm CMB,0}=2.7260 \pm 0.0013~{\rm K}$ (i.e. a precision of $\sim0.05$ percent), is based only on the Far-InfraRed Absolute Spectrophotometer (FIRAS) experiment, recalibrated to the Wilkinson Microwave Anisotropy Probe (WMAP) data \citep{Fixsen2009}.

The temperature of the CMB can also be inferred from the relative populations of the lowest rotational states of some molecular species that reside in the diffuse molecular medium of galaxies. In fact, the CMB was first detected by \citet[][see also, \citealt{Adams1941}]{McKellar1941} using this technique. Even though the meaning of the estimated excitation temperature was not fully appreciated at the time, the CMB temperature inferred by \citet{McKellar1941} was more accurate than the direct measurement by \citet{PenziasWilson1965}. The best molecular CMB thermometer is the cyano radical (CN), which has an energy difference of $\sim4.69\times10^{-4}~{\rm eV}$ ($\equiv2.6414$\,mm) between the two lowest rotational levels of CN; this wavelength is near the peak of the CMB blackbody curve ($\lambda_{\rm peak}\simeq1.06\,{\rm mm}$). The first rotationally exicted state of CN is therefore largely dominated by the absorption of 2.6414\,mm CMB photons.

Strictly speaking, the excitation temperature provides an upper limit on the CMB temperature, since local sources can also contribute to the excitation of CN, particularly electron impacts \citep{Thaddeus1972,BlackVanDishoeck1991}. In principle, the local excitation contribution to the observed level populations can be estimated empirically by observing 2.6414\,mm emission along the line-of-sight \citep{Penzias1972,Crane1989,Palazzi1990,Palazzi1992,Roth1993}. This correction is primarily affected by the uniformity of the emission; if the source of CN emission does not uniformly fill the antenna beam, the local source correction would be underestimated. Current data indicate that local sources may contribute to CN excitation at the level of $0.01-0.1~{\rm K}$, depending on the source. Also note that there are no physical processes that can drive the CN level populations to an excitation temperature that is below the CMB temperature. Therefore, sightlines with a CN excitation temperature below the CMB temperature are likely affected by systematic biases.

Despite the possibility that some sightlines might experience some heating due to local excitation, interstellar CN absorption offers a highly complementary probe of the CMB temperature beyond the local environment of Earth. Around the time of the Cosmic Background Explorer (COBE) mission \citep{Mather1990}, the CMB temperatures inferred by CN excitation had a typical precision of $\sim1.5$ percent ($\equiv\Delta T\simeq0.04~{\rm K}$; \citealt{MeyerJura1985,KaiserWright1990,Roth1995}). These measures are still in excellent agreement with the currently accepted CMB temperature reported by \citet{Fixsen2009}. On the other hand, a single high precision measurement of the CN excitation temperature along the line of sight to $\zeta$~Oph ($T_{\rm ex}\simeq2.796^{+0.014}_{-0.039}~{\rm K}$; \citealt{Crane1989}) is slightly elevated relative to the CMB. Based on a sample of 34 CN absorbers, \citet{Palazzi1992} reported a weighted mean value of $T_{\rm ex}\simeq2.817\pm0.022~{\rm K}$, which is in good agreement with the CN absorber towards $\zeta$~Oph. Both of these values are modestly higher than the values reported by other authors, indicating an excess of 0.091~K above the \citet{Fixsen2009} result. Some of this discrepancy could be explained by the multi-component structure of the CN absorber towards $\zeta$~Oph, that only became apparent with data of especially high spectral resolution ($R>600\,000$; \citealt{Lambert1990,Crawford1994}). This  highlighted the importance of using only the highest quality data --- both in terms of spectral resolution and S/N ratio --- when determining the excitation temperature of CN absorbers.

Using a sample of 13 sightlines, \citet{Ritchey2011} analysed data of extremely high S/N ($\gtrsim1\,000$) and spectral resolution ($R\sim100\,000-200\,000$). Based on these data, \citet{Ritchey2011} reported a weighted average excitation temperature of $T_{\rm ex}\simeq2.754\pm0.002~{\rm K}$, which exceeds the CMB temperature by $\sim0.035$K. \citet{Slyk2008} reported CN excitation temperatures along 73 sightlines based on data of ${\rm S/N}=80-200$ and somewhat lower spectral resolution ($R\lesssim120,000$). These authors reported a much larger 0.58~K average excess over the CMB temperature. The significant difference between these two results can be attributed to the adopted Doppler parameter of the absorption lines. \citet{Slyk2008} assumed a Doppler parameter $b=1~{\rm km~s}^{-1}$ for all absorbers, which is systematically higher than that found by other studies. Since the strongest CN line is often saturated for the strongest CN absorbers, an over-estimated Doppler parameter leads to an under-estimated column density and therefore an over-estimated excitation temperature. Similarly, the \citet{Palazzi1992} Doppler parameters are somewhat elevated compared to the \citet{Ritchey2011} values, and this can explain the difference between the excitation temperatures reported by these authors. Furthermore, in some of the aforementioned studies, the intrinsic dispersion among the reported excitation temperatures exceeds the quoted errors. This suggests that the miscalculated Doppler parameters primarily act as an unaccounted for systematic uncertainty.

The key takeaway from this brief history is that the CN absorption lines are intrinsically very narrow, and likely unresolved when the instrument resolution $R\lesssim200\,000$, equivalent to a full-width at half maximum (FWHM) velocity, $V_{\rm FWHM}\gtrsim1.5~{\rm km~s}^{-1}$. Unresolved and saturated line profiles lead to biased Doppler parameters and hence biased column densities. For example, four of the thirteen absorbers reported by \citet{Ritchey2011} exhibit excitation temperatures significantly ($>4\sigma$) below the CMB temperature, which they attribute to biases in the inferred Doppler parameters; \citet{Ritchey2011} conclude that these four absorbers can be brought into agreement with the CMB temperature if the inferred Doppler parameters were increased by just $0.06~{\rm km~s}^{-1}$. At present, using CN absorbers to infer the present day CMB temperature is therefore limited by systematic uncertainties in the profile modelling. Nevertheless, the Milky Way sightlines discovered to date offer the perfect laboratory to study the limitations of measuring the CMB temperature from the relative level populations of some molecules.

At present, electronic CN absorption line systems have been identified in the Milky Way interstellar medium, along several sightlines to the Magellanic Clouds \citep{Welty2006}, and in the host galaxies of several nearby supernovae \citep{Patat2007,CoxPatat2014,Ritchey2015a}. Given the sensitivity of the CN level populations to the CMB temperature, one of the key goals of future telescope facilities is to detect CN absorption in high redshift galaxies in order to determine the redshift evolution of the CMB temperature \citep{MartinsCooke2024}. Molecular CO absorption lines have been identified along the lines of sight through several high redshift galaxies \citep{Srianand2008,Noterdaeme2011,Klimenko2020}, allowing the CMB temperature to be inferred at redshifts $1<z<3$ with a precision of $\sim10$ percent. At $z\lesssim1$, the only molecular absorption system that has been used to infer the CMB temperature is toward the gravitationally lensed quasar PKS\,1830$-$211 \citep{Muller2013}, yielding a two percent determination of the CMB temperature at $z\simeq0.89$ based on a multi-transition excitation analysis. Recently, a $z=6.34$ water absorption line system imprinted on the CMB was discovered, yielding the highest redshift determination of the CMB temperature \citep{Riechers2022}, albeit with a $1\sigma$ confidence interval spanning $\sim14\,{\rm K}$. The CMB temperature can also be inferred from inverse Compton scattering of CMB photons off hot intracluster gas \citep{Battistelli2002}. The distortions seen in CMB maps allows the CMB temperature to be measured at the redshift of the hot gas with high precision on individual systems ($\sim5$ percent; \citealt{Hurier2014,Luzzi2015,deMartino2015,Li2021}), and now with large samples ($\gtrsim100$ systems). However, such measures are limited to $z<1$ due to the scarcity of high redshift clusters. In all environments found so far, the CMB temperature does not deviate from the standard cosmological relationship, $T_{\rm CMB}(z)=T_{\rm CMB,0}\,(1+z)$, which is based on the Friedmann–Lema\^itre–Robertson–Walker metric. Deviations from this simple relationship would indicate new physics beyond the Standard Model (see, e.g. \citealt{MartinsVieira2024}), and perhaps the most promising approach to precisely pin down this relationship in the future is with electronic CN absorption in high redshift galaxies.

With a view to these future measures of CN absorption, in this paper we reassess the potential systematic uncertainties of CN absorption using the highest quality data currently available of Milky Way sightlines. We assess the precision that can be reached with current instrumentation, and discuss the leading systematic uncertainties. In Section~\ref{sec:data}, we discuss the archival data used in this work, and the approach that we adopted to reduce the data. We compile a list of the relevant molecular data and summarise the theory of CN rotational excitation in Section~\ref{sec:cn}, before outlining our analysis strategy in Section~\ref{sec:analysis}. In Section~\ref{sec:discussion}, we report a new measurement of the excitation temperature, and the intrinsic (systematic) scatter of the measurements. Finally, we summarise our main conclusions in Section~\ref{sec:conc}.

\section{Observations and data reduction}
\label{sec:data}

\begin{table*}
    \centering
    \caption{Journal of observational data that are used in this study}
    \label{tab:obs}
    \begin{tabular}{lccccccc}
        \hline
        Star      & UT date      & Instrument & Wavelength  & Exposure & $V_{\rm FWHM}\,^{a}$      & S/N pixel$^{-1}$ & pixel size\\
                  & (yyyy-mm-dd) &            & range (\AA) & time (s) & $({\rm km~s}^{-1})$ & at 3875\,\AA & ${\rm km~s}^{-1}$\\
        \hline
        HD 24534  & 2018-11-16 & VLT/ESPRESSO & $3783-7908$ & 1000 & $1.398\pm0.005$, $1.456\pm0.009$ & 115 & 0.5 \\
        HD 27778  & 2018-11-09 & VLT/ESPRESSO & $3783-7908$ & 980  & $1.398\pm0.005$, $1.456\pm0.009$ & 171 & 0.5 \\
        HD 62542  & 2018-11-30 & VLT/ESPRESSO & $3783-7908$ & 1700 & $1.398\pm0.005$, $1.456\pm0.009$ & 193 & 0.5 \\
        HD 73882  & 2018-11-09 & VLT/ESPRESSO & $3783-7908$ & 570  & $1.393\pm0.007$, $1.465\pm0.006$ & 90 & 0.5 \\
        HD 147933 & 2019-03-02 & VLT/ESPRESSO & $3783-7908$ & 110  & $1.394\pm0.003$, $1.466\pm0.010$ & 154 & 0.5 \\
        HD 149757 & 1993-05-11 & AAT/UHRF     & $3875.52-3877.78$ & 4073 & $0.576\pm0.007$ & 150 & 0.15 \\
        HD 152236 & 1993-05-11 & AAT/UHRF & $3875.56-3877.34$ & 1994 & $0.579\pm0.003$ & 54 & 0.15 \\
                  & 1994-04-29 & AAT/UHRF & $3875.47-3877.57$ & 3852 & $0.568\pm0.002$ & 64 & 0.15 \\
                  & 2019-03-03 & VLT/ESPRESSO & $3783-7908$ & 74 & $1.394\pm0.003$, $1.466\pm0.010$ & 89 & 0.5 \\
        HD 152270 & 1994-04-24, 1994-04-29 & AAT/UHRF & $3875.47-3877.94$ & 8283 & $0.563\pm0.003$ & 30 & 0.15 \\
        \hline
    \end{tabular}
    $^{a}$ For the ESPRESSO data, we list the FWHM values of each spectrum separately.
\end{table*}

Given the difficulty of accurately measuring Doppler parameters of unresolved lines, we restrict our sample to include all CN absorbers that have been observed with the highest possible spectral resolution ($R\gtrsim500\,000$), in addition to all observations of CN absorbers made with the ultra-stable high resolution ESPRESSO\footnote{Echelle SPectrograph for Rocky Exoplanets and Stable Spectroscopic Observations} spectrograph ($R\simeq190\,000$; \citealt{Pepe2021}) at the Very large Telescope (VLT) facility. This selection resulted in a total of eight sightlines. Our goal is to use these high quality data to accurately pin down the cloud model, and minimise systematic uncertainties.
A summary of the observations used in our analysis is provided in Table~\ref{tab:obs}, and in the following subsections, we describe the data reduction strategy used for the data acquired with each of these instruments. We note that many of our targets have also been observed with somewhat lower resolution ($R\sim40,000-100,000$) echelle spectrographs, including the Ultraviolet and Visual Echelle Spectrograph (UVES) on the VLT. These data were not included in the present study, since the UVES instrument resolution varies considerably from order-to-order and exposure-to-exposure.

\subsection{AAT UHRF}
The Anglo-Australian Telescope (AAT) Ultra High Resolution Facility (UHRF) is a decommissioned spectrograph that operated during the years 1993$-$2013 \citep{Diego1995}.\footnote{Although, a prototype was swiftly built a few years earlier to observe the supernova SN\,1987A \citep{Pettini1988}.} The only CN absorbers observed with UHRF during this time were HD\,149757 ($\equiv\zeta$\,Oph), HD\,152236 ($\equiv\zeta^{1}~{\rm Sco}$), HD\,152270, and HD\,169454. For details of the observations, refer to the series of papers by \citet{Crawford1994,Crawford1995,Crawford1997}. Despite being three decades old, these data are the highest spectral resolution observations of CN absorbers currently available. We retrieved the UHRF data from the AAT data archive.\footnote{The data archive can be accessed from the following link:\\ \url{https://archives.datacentral.org.au/query}}

We have added support for UHRF data reduction within the \pypeit\ data reduction software \citep{Prochaska2020}, but we note that several non-standard steps were required to optimally extract the data. Our modifications were implemented on a development version that postdates version 1.15.0 of \pypeit. We subtracted the bias level based on the overscan regions of the detector. Since there were no flatfield exposures available in the archive, we did not correct for pixel-to-pixel sensitivity variations, nor did we perform a spatial illumination correction; this should not affect the quality of the data reduction, since the spatial object profile is typically projected onto $\gtrsim50$ detector pixels at a given wavelength. We manually map the wavelength solution of each exposure, based on $\sim6$ ThAr lines, with typical root-mean-square (RMS) residuals of $\lesssim0.05$\,pixels. This map allows us to account for the wavelength tilt along the spectral direction of the detector. We iteratively mask cosmic rays in individual exposures during the optimal extraction. The object profiles are quite extended, due to the image slicer, and vary with wavelength. The standard \pypeit\ algorithm for fitting the object spatial profile does not perform well in these circumstances. We therefore manually defined the `sky' regions, and fit the object profile using Gaussian kernel density estimation with a bandwidth of $1-2$ pixels. This was allowed to smoothly vary as a function of wavelength, and produced residuals that are close to the Poisson limit. As a final step, we manually combined the data with \texttt{UVES\_popler} (version 1.05)\footnote{\texttt{UVES\_popler} is available from the following link:\\
\url{https://github.com/MTMurphy77/UVES_popler/}} \citep{Murphy2019} to mask bad pixels and cosmic rays, and sample the final combined spectrum with a pixel size of $0.15~{\rm km~s}^{-1}$. Unfortunately, we note that the data towards HD\,169454 are extremely low S/N per detector pixel, highly saturated, and did not provide a meaningful measurement of the CN absorption. We therefore do not consider this system further, given the goals of the present study.

\subsection{VLT ESPRESSO}

ESPRESSO is a recently commissioned ultra-stable fibre-fed high resolution spectrograph that is located at the incoherent combined Coud{\'e} facility of the VLT. ESPRESSO can be fed by any one of the 8\,m VLT unit telescopes, or simultaneously be fed by all four 8\,m telescopes. The former mode offers the highest nominal resolution ($R\simeq190\,000$). One of the primary goals of ESPRESSO is to reach a wavelength calibration stability of $10~{\rm cm~s}^{-1}$ over a time span of 10 years. Towards this goal, the instrument line spread function (LSF) of ESPRESSO has recently been mapped using a laser frequency comb (LFC), revealing that the LSF varies across the detector and exhibits substantial asymmetries \citep{Schmidt2024} that need to be accounted for when high precision radial velocities are desired. Furthermore, a preliminary investigation by this study found that the ESPRESSO LSF varies slightly as a function of time. For the purposes of this work, the LSF plays a critical role in determining the intrinsic Doppler widths of the CN absorption lines, and we discuss our approach to dealing with this issue in Section~\ref{sec:analysis}.

We searched the ESO data archive\footnote{The data archive can be accessed from the following link:\\ \url{https://archive.eso.org/wdb/wdb/eso/espresso/form}} for any known CN absorbers that have been observed with the highest resolution mode of ESPRESSO. A total of six stars were returned, including one star (HD\,152236) that has also been observed with UHRF (for further details, see Table~\ref{tab:obs}). All six stars considered here were observed as part of the \citet{DeCia2021} survey (Programme ID: 0102.C-0699(A)). We used \texttt{ESOREX} version 3.13.7 to reduce the ESPRESSO data. The organisation of raw frames and data reduction steps were executed using a script written by Jens Hoeijmakers\footnote{The script is available from:\\
\url{https://github.com/Hoeijmakers/ESPRESSO_pipeline}} with minor modifications. As part of a further post-processing step, we found that the sky subtraction created spurious noise in our high S/N spectra. We therefore applied a Savitzky–Golay filter of order 2 and window length 50 to the sky spectra before subtracting the result from the science frames. We visually inspected the multiple spectra for consistency, then identified and masked several groups of warm pixels (see \citealt{PasquiniMilakovic2024}) before our analysis. These warm pixels are revealed as significant deviations when comparing multiple spectra that fall on different parts of the detector. We neither combine nor resample the ESPRESSO spectra; instead we opted to analyse the data with the native pixel scale, corresponding to a pixel size of $\sim0.5~{\rm km~s}^{-1}$. By not combining the data, we can analyse each ESPRESSO spectrum separately with its own LSF. Each ESPRESSO spectrum can also be analysed independently, to check if the derived model parameters of the two independent spectra mutually agree with the other.

\section{CN as a thermometer}
\label{sec:cn}

\begin{table}
    \centering
    \caption{Interstellar $^{12}$C$^{14}$N molecular data \citep{Brooke2014}}
    \label{tab:molecular}
    \begin{tabular}{lcccc}
        \hline
        Line & $\lambda_{\rm 0,vac}$ & $\lambda_{\rm 0,air}$ & $f$ & $\Gamma$\\
            & (\AA)     & (\AA)     &     &  ($10^{7}~{\rm s}^{-1}$)   \\
        \hline
                $R_{1}$(2)  & $3874.46031$ & $3873.36232$ & $0.01927$ & $1.495$ \\
                $R_{2}$(2)  & $3874.46660$ & $3873.36860$ & $0.02023$ & $1.495$ \\
        $^{\rm R}Q_{21}$(2) & $3874.46932$ & $3873.37132$ & $0.00096$ & $1.495$ \\
                $R_{1}$(1)  & $3875.08975$ & $3873.99159$ & $0.02023$ & $1.495$ \\
                $R_{2}$(1)  & $3875.09456$ & $3873.99640$ & $0.02248$ & $1.495$ \\
        $^{\rm R}Q_{21}$(1) & $3875.09619$ & $3873.99803$ & $0.00225$ & $1.495$ \\
                $R_{1}$(0)  & $3875.69914$ & $3874.60082$ & $0.02247$ & $1.495$ \\
        $^{\rm R}Q_{21}$(0) & $3875.70300$ & $3874.60468$ & $0.01124$ & $1.495$ \\
        $^{\rm P}Q_{12}$(1) & $3876.85624$ & $3875.75762$ & $0.01123$ & $1.495$ \\
                $P_{1}$(1)  & $3876.85787$ & $3875.75926$ & $0.01123$ & $1.495$ \\
        $^{\rm P}Q_{12}$(2) & $3877.40258$ & $3876.30382$ & $0.00225$ & $1.495$ \\
                $P_{1}$(2)  & $3877.40532$ & $3876.30655$ & $0.01347$ & $1.495$ \\
                $P_{2}$(2)  & $3877.40646$ & $3876.30770$ & $0.01123$ & $1.495$ \\
        \hline
    \end{tabular}

\end{table}

\begin{table}
    \centering
    \caption{Interstellar $^{13}$C$^{14}$N molecular data \citep{Sneden2014}}
    \label{tab:molecular1314}
    \begin{tabular}{lcccc}
        \hline
        Line & $\lambda_{\rm 0,vac}$ & $\lambda_{\rm 0,air}$ & $f$ & $\Gamma$\\
            & (\AA)     & (\AA)     &     &  ($10^{7}~{\rm s}^{-1}$)   \\
        \hline
                $R_{1}$(2)  & $3874.67940$ & $3873.58135$ & $0.01924$ & $1.493$ \\
                $R_{2}$(2)  & $3874.68528$ & $3873.58723$ & $0.02020$ & $1.493$ \\
        $^{\rm R}Q_{21}$(2) & $3874.68791$ & $3873.58986$ & $0.00096$ & $1.493$ \\
                $R_{1}$(1)  & $3875.28280$ & $3874.18459$ & $0.02020$ & $1.493$ \\
                $R_{2}$(1)  & $3875.28732$ & $3874.18911$ & $0.02244$ & $1.493$ \\
        $^{\rm R}Q_{21}$(1) & $3875.28888$ & $3874.19067$ & $0.00224$ & $1.493$ \\
                $R_{1}$(0)  & $3875.86696$ & $3874.76860$ & $0.02244$ & $1.493$ \\
        $^{\rm R}Q_{21}$(0) & $3875.87061$ & $3874.77225$ & $0.01122$ & $1.493$ \\
        $^{\rm P}Q_{12}$(1) & $3876.97615$ & $3875.87750$ & $0.01121$ & $1.493$ \\
                $P_{1}$(1)  & $3876.97773$ & $3875.87908$ & $0.01121$ & $1.493$ \\
        $^{\rm P}Q_{12}$(2) & $3877.49990$ & $3876.40111$ & $0.00224$ & $1.493$ \\
                $P_{1}$(2)  & $3877.50251$ & $3876.40373$ & $0.01345$ & $1.493$ \\
                $P_{2}$(2)  & $3877.50355$ & $3876.40476$ & $0.01121$ & $1.493$ \\
        \hline
    \end{tabular}

\end{table}

\begin{table}
    \centering
    \caption{Interstellar $^{12}$C$^{15}$N molecular data \citep{Sneden2014}}
    \label{tab:molecular1215}
    \begin{tabular}{lcccc}
        \hline
        Line & $\lambda_{\rm 0,vac}$ & $\lambda_{\rm 0,air}$ & $f$ & $\Gamma$\\
            & (\AA)     & (\AA)     &     &  ($10^{7}~{\rm s}^{-1}$)   \\
        \hline
                $R_{1}$(2)  & $3874.61969$ & $3873.52166$ & $0.01925$ & $1.494$ \\
                $R_{2}$(2)  & $3874.62525$ & $3873.52721$ & $0.02021$ & $1.494$ \\
        $^{\rm R}Q_{21}$(2) & $3874.62788$ & $3873.52984$ & $0.00096$ & $1.494$ \\
                $R_{1}$(1)  & $3875.22984$ & $3874.13165$ & $0.02021$ & $1.494$ \\
                $R_{2}$(1)  & $3875.23412$ & $3874.13593$ & $0.02245$ & $1.494$ \\
        $^{\rm R}Q_{21}$(1) & $3875.23569$ & $3874.13749$ & $0.00224$ & $1.494$ \\
                $R_{1}$(0)  & $3875.82054$ & $3874.72219$ & $0.02245$ & $1.494$ \\
        $^{\rm R}Q_{21}$(0) & $3875.82405$ & $3874.72570$ & $0.01122$ & $1.494$ \\
        $^{\rm P}Q_{12}$(1) & $3876.94224$ & $3875.84360$ & $0.01122$ & $1.493$ \\
                $P_{1}$(1)  & $3876.94380$ & $3875.84516$ & $0.01122$ & $1.493$ \\
        $^{\rm P}Q_{12}$(2) & $3877.47198$ & $3876.37320$ & $0.00224$ & $1.494$ \\
                $P_{1}$(2)  & $3877.47461$ & $3876.37583$ & $0.01346$ & $1.494$ \\
                $P_{2}$(2)  & $3877.47550$ & $3876.37672$ & $0.01122$ & $1.494$ \\
        \hline
    \end{tabular}

\end{table}

\subsection{Molecular data}
\label{sec:moledata}

In this paper, we focus on the CN \band\ (0,0) vibronic band near $3875\,$\AA. The strongest transitions are commonly referred to as $R(0)$, $R(1)$, $R(2)$, $P(1)$, $P(2)$ for convenience, but each of these lines is in fact a blend of several transitions (see \citealt{Federman1984} for a useful summary). For example, the $R(0)$ line is a blend of the $R_{1}$(0) and $^{\rm R}Q_{21}$(0) transitions, while the $R(1)$ line is a blend of the $R_{1}$(1), $R_{2}$(1), and $^{\rm R}Q_{21}$(1) lines. Given the high spectral resolution of the data that we analyse in this work and the typical separations of these lines ($\sim0.1-0.5\,{\rm km~s}^{-1}$), it is important to properly account for each transition as part of our analysis. We adopt the \citet{Brooke2014} $^{12}$C$^{14}$N molecular data, which are reproduced in Table~\ref{tab:molecular} for convenience. For $^{13}$C$^{14}$N and $^{12}$C$^{15}$N, we adopt the \citet{Sneden2014} molecular data, which are provided in Table~\ref{tab:molecular1314} and ~\ref{tab:molecular1215}, respectively. We include both vacuum ($\lambda_{\rm 0,vac}$) and air ($\lambda_{\rm 0,air}$) wavelengths for completeness; our analysis uses vacuum wavelengths.

\subsection{Rotational excitation}

If we assume the CN molecules are in local thermodynamic equilibrium with a blackbody radiation field of temperature $T_{lu}$, then the relative population of a lower rotational level, $l$, and an upper rotational level, $u=l+1$, is described by the Boltzmann equation:
\begin{equation}
    \label{eqn:lte}
    \frac{N[N''=u]}{N[N''=l]} = \frac{g_{u}}{g_{l}}\,\exp\Big[-\frac{h\nu_{lu}}{k_{\rm B}T_{lu}}\Big]
\end{equation}
where $N$ is the column density of the $N''$ quantum level, $g_{l}=\sum(2J_{l}+1)$ and $g_{u}=\sum(2J_{u}+1)$ are the statistical weights of the rotational levels, and $\nu_{lu}$ is the transition frequency. The column density of the $N''=1$ level is given by
\begin{equation}
    N[N''=1] = N[N''=1, J''=1/2] + N[N''=1, J''=3/2].
\end{equation}
where the column density arising from a specific $J$ level is explicitly indicated. This distinction is necessary, because the $J=1/2$ and $J=3/2$ levels have a different $\nu_{lu}$ frequency in Equation~\ref{eqn:lte}. In previous studies, this small effect has not been included; it only impacts the derived temperature at the $\sim0.2$ percent level ($\simeq0.005\,{\rm K}$). However, this level of precision is comparable to the uncertainties reported for several previously measured CN excitation temperatures. The relative population of the $N''=1$ and $N''=0$ levels is therefore,
\begin{equation}
    \label{eqn:r1r0}
    \frac{N[N''=1]}{N[N''=0]} = \exp\Big[-\frac{h\nu_{01}}{k_{\rm B}T_{\rm 01}}\Big]\bigg(1+2\exp\Big[-\frac{h\Delta\nu_{1}}{k_{\rm B}T_{\rm 01}}\Big]\bigg)
\end{equation}
where transitions of $^{12}{\rm C}^{14}{\rm N}$ from the $J=1/2$ level have an energy separation $h\nu/k_{\rm B}=5.4312\,{\rm K}$, while those from $J=3/2$ have an energy separation $h\nu/k_{\rm B}=5.4469\,{\rm K}$. Therefore, $h\nu_{01}/k_{\rm B}=5.4312~{\rm K}$ and $h\Delta\nu_{1}/k_{\rm B}=(5.4469-5.4312)~{\rm K}=0.0157~{\rm K}$.
The corresponding values for $^{13}{\rm C}^{14}{\rm N}$ are $h\nu_{01}/k_{\rm B}=5.2061~{\rm K}$ and $h\Delta\nu_{1}/k_{\rm B}=(5.2210-5.2061)~{\rm K}=0.0149~{\rm K}$. Similarly, for $^{12}{\rm C}^{15}{\rm N}$ the values are $h\nu_{01}/k_{\rm B}=5.2652~{\rm K}$ and $h\Delta\nu_{1}/k_{\rm B}=(5.2802-5.2652)~{\rm K}=0.0150~{\rm K}$.

Note, the temperature ($T_{01}$) that appears in Equation~\ref{eqn:r1r0} is referred to as the excitation temperature, and only reflects the blackbody temperature in the limit that local sources of excitation (e.g. collisions with electrons) are negligible. A small correction to the excitation temperature due to local sources ($\Delta T_{\rm loc}$) allows one to recover the CMB temperature, i.e. $T_{\rm CMB}=T_{lu}-\Delta T_{\rm loc}$. A similar approach can be carried out for the relative population of the $N''=2$ and $N''=1$ levels. However, we point out that the excitation temperature derived from the first and second rotationally excited levels ($T_{12}$) is usually not as precisely measured as $T_{01}$. We therefore focus our attention in this paper on obtaining new precise measures of $T_{01}$. For sightlines with transitions detected from the $N''=2$ level (i.e. either the $R(2)$ or $P(2)$ lines), we adopt the following equation for inferring the excitation temperature of the first and second rotationally excited levels:
\begin{equation}
    \label{eqn:r2r1}
    \frac{N[N''=2]}{N[N''=1]} = \frac{5}{3}\exp\Big[-\frac{h\nu_{12}}{k_{\rm B}T_{\rm 12}}\Big]
\end{equation}
where $h\nu_{12}/k_{\rm B}=10.883~{\rm K}$. Hereafter, we do not explicitly mention the $N''$ levels when quoting column densities. Instead, we use the notation $N(1)/N(0)$ to represent $N\!(N''\!=\!1)/N\!(N''\!=\!0)$, and we adopt $N(2)/N(1)$ to represent $N\!(N''\!=\!2)/N\!(N''\!=\!1)$.

\section{Analysis}
\label{sec:analysis}

Our analysis offers four key improvements over previous work on CN excitation:
(1) we use an absorption line profile fitting software that has not previously been applied to the analysis of CN (see Section~\ref{sec:profilefitting} for details);
(2) we have attempted to identify the leading causes of systematic uncertainty, and included these effects in the profile analysis. This ensures that the final parameter uncertainties include the dominant systematic uncertainties;
(3) we have constructed a model of the instrumental line spread function of each exposure; and
(4) we have performed a blind analysis to limit the possibility of human bias.
In this section, we highlight the key aspects of our analysis.

\subsection{Line spread function}
\label{sec:lsf}

The total width of the observed absorption lines contains a contribution of intrinsic broadening and the instrument broadening. The intrinsic broadening includes a turbulent (i.e. macroscopic) term as well as a thermal (i.e. microscopic) term, while the instrument broadening depends on various properties of the instrument (e.g. slit width, wavelength, detector pixel position, time of observation). The instrument broadening is modelled with a line spread function (LSF); most absorption line analyses usually assume the LSF is a Gaussian profile with a nominal FWHM that is derived from unresolved line profiles of a comparison lamp. In reality, the LSF is asymmetric and wavelength dependent. In this work, we aim to provide a proper accounting of the LSF shape and its associated uncertainty to determine the impact of the LSF on the derived excitation temperatures. We deem this to be an important part of the analysis, since underestimating the FWHM directly translates to overestimating the intrinsic Doppler parameter. For strong absorption lines, this causes the column density of the stronger $R(0)$ line to be underestimated relative to the column densities of the weaker $R(1)$ and $P(1)$ lines. If unaccounted for, this systematic column density bias introduces a biased (in this example, overestimated) excitation temperature. The importance of an accurate cloud model, including the effects due to saturation and a poorly known LSF, has been discussed by several authors in the past \citep[e.g.][]{Palazzi1992,Roth1993,Roth1995,Slyk2008,Ritchey2011}.

For most of the sightlines analysed in this work, the CN absorption lines are either unresolved or only marginally resolved, even at the spectral resolution of the UHRF instrument. We therefore develop a strategy to account for the unknown shape of the LSF. We model the data with two different choices of the LSF that should represent the `extremes' of the intrinsic LSF widths. With the first approach, we model the LSF as a multi-component Gaussian (typically with $N_{\rm gauss}=4$ components), of the form:
\begin{equation}
\label{eqn:lsf}
    LSF(\lambda)_{\rm G} = \sum_{i=1}^{N_{\rm gauss}} \!\!\!a_{i}\,\exp\bigg[-\frac{(\lambda-\lambda_{0}-\delta_{i})^{2}}{2\sigma_{i}^{2}}\bigg]
\end{equation}
where all of the parameters with subscripts are free model parameters, with the exception of $a_{1}=1$ and $\delta_{1}=0$. Note that $0 \leq a_{i} \leq 1$, and we normalise the total profile so that $\int LSF(\lambda)_{\rm G}\,{\rm d}\lambda=1$. We perform a joint fit to all ThAr emission lines in the wavelength interval $3873-3878$\,\AA\ (there are five ThAr emission lines that we include in the fit). Typically, our fit includes $\pm20$ pixels around the centre of each line. In addition to the free parameters in Equation~\ref{eqn:lsf}, we include three model parameters for each ThAr emission line to fit the local continuum level (one parameter), the centroid of the line profile (one parameter) and the amplitude of the ThAr emission line (one parameter). We consider this approach to provide an estimate of the \emph{broadest} possible LSF allowed by the data. Specifically, if the ThAr lines are completely unresolved and the illumination of the slicer is identical for both the science target and the calibration lamp, then this LSF should accurately represent the LSF of the science observations. This assumption is widely adopted as the `standard' approach.

The second approach that we consider aims to empirically construct a model of the \emph{narrowest} possible LSF. We do this as part of a two stage process. During the first stage, we fit the ThAr data using a multi-component Voigt profile, equivalent to:
\begin{equation}
    LSF(\lambda,\gamma)_{\rm V} = \!\!\!\sum_{i=1}^{N_{\rm gauss}} \!\!\!a_{i}\,\exp\bigg[-\frac{(\lambda-\lambda_{0}-\delta_{i})^{2}}{2\sigma_{i}^{2}}\bigg] * \frac{\gamma}{\gamma^{2} + (\lambda-\lambda_{0}-\delta_{i})^{2}}
\end{equation}
where $\gamma$ is a free parameter, and is the same for all components, and we fix $a_{1}=1$ and $\delta_{1}=0$. This functional form is adopted to account for the intrinsic natural broadening of the ThAr emission lines. Once the model parameters have been optimised, we then set the damping constant, $\gamma=0$, and normalise the profile. This second step assumes that the wings of the ThAr emission lines are purely due to natural broadening. By setting $\gamma=0$, we are removing the contribution of the (assumed) intrinsic ThAr line width to obtain a measure of the instrumental broadening alone. This of course assumes that the ThAr wings are intrinsic to the lamp and not the spectrograph. This approach therefore underestimates the true width of the LSF, and provides an estimate of the \emph{narrowest} possible LSF allowed by the data.
Taken together, these two `extreme' LSF models are used to infer the range of plausible column density ratios (for further details, see Section~\ref{sec:discussion}). To estimate the uncertainty of each LSF, we perform a bootstrap analysis to generate 20 realisations of each LSF for every target. We use these 20 bootstrap samples in conjunction with the line profile fitting to obtain an estimate of the systematic uncertainty of each LSF (see Section~\ref{sec:profilefitting}).

For the UHRF data, we extracted a spectrum of the ThAr comparison lamp using the same optimal profile as the science target. We note that modelling the LSF based on the ThAr lamp assumes that the illumination of the entrance slicer is the same for both the ThAr lamp frame and the science target. We consider this to be a reasonable assumption, given that the slice width of the UHRF is $\sim0.05''$, which is much smaller than the typical seeing conditions at Siding Spring \citep{Goodwin2013}.\footnote{We have no record of the seeing conditions during the observations, but the typical seeing at the AAT is $1.2''$. We also note the seeing is very rarely less than $0.5''$, which is an order of magnitude larger than the slice width.} The wavelength coverage of UHRF typically includes $\sim5$ ThAr emission lines; all of these line profiles are mutually consistent with one another, and we highlight that the LSF is asymmetric for all observations reported here. We further note that the UHRF LSF of HD\,152236 taken in 1993 is sufficiently different from the 1994 data that we decided not to combine these data; instead, we model each spectrum separately with their own asymmetric LSF.

We use the same approach described above to model the ESPRESSO LSF. The ESPRESSO data also have an additional advantage over other spectrographs; because the ESPRESSO optical design includes a pupil slicer \citep{Riva2014}, each fibre is imaged on the detector twice, and each of these spectra can have a different LSF \citep{Schmidt2024}. We also note that ESPRESSO is equipped with a double scrambler that ensures the illumination of the spectrograph is homogeneous and stable \citep{Pepe2021}; each slice should provide a nearly identical spectrum of the same source, acquired at the same time. This offers an excellent cross-check of the analysis, since both fibres should produce an identical result; if the analysis reveals a difference between these separate spectra, then it could indicate an issue with either the data reduction or the LSF determination. As mentioned above, ESPRESSO is also equipped with a LFC that covers the wavelength range $4300-7900$\,\AA\ (see \citealt{Schmidt2024}). We cross-checked our LSF fitting analysis with observations of LFC and ThAr calibrations that are available in the ESO archive, and found that the ThAr emission lines are in fact \emph{narrower} than the LFC lines. We did not exhaustively check this is the case over the entire ESPRESSO wavelength range, but this may indicate that the LFC linewidth is more resolved than the ThAr lines. Since this is not relevant to our study, we do not consider it further. As a final note, we find that the ESPRESSO data are often well fit near 3875\AA\ by a single Voigt profile using the approach described above. For reference, the measured FWHMs and the associated uncertainties of all spectra are listed in Table~\ref{tab:obs}. For ESPRESSO, we list the FWHM of each slice separately.

\subsection{Line profile fitting}
\label{sec:profilefitting}

We have used a development version of the Absorption LIne Software (\textsc{alis}) package\footnote{\textsc{alis} is available from the following repository:\\ \url{https://github.com/rcooke-ast/ALIS}} \citep{Cooke2014}. This software was originally designed to measure the deuterium abundance of near-pristine absorption line systems, and has been adapted for the purposes of this study. \textsc{alis} uses the Levenberg–Marquardt algorithm to perform a non-linear least-squares fit to the data given a set of model parameters \citep{Markwardt2009}. The full model is defined by a continuum, a series of absorption lines, and a line-spread function (see Section~\ref{sec:lsf} and below). We note that all parameters of the model are fit simultaneously, so that the errors on the final derived column density ratios include the dominant uncertainties associated with the modelling procedure.

For the continuum, we use a low-order Legendre polynomial (typically of order 3) in the vicinity of each absorption line. Each absorption line is modelled as a Voigt profile, which consists of three parameters: a column density, a velocity shift relative to the barycentre of the Solar System, and a Doppler parameter. All absorption lines of a given component are assumed to have the same Doppler parameter. We assume that thermal broadening is negligible, and model the absorption lines with macroscopic turbulence only. This choice does not impact our inferred excitation temperatures, since the excitation temperatures are based on the ratios of absorption lines that have the same molecular mass. Therefore, turbulent and thermal broadening are degenerate. Each pixel is sampled by 20 `sub-pixels'; the model is evaluated on this fine wavelength grid, and then resampled to the native pixel scale.

\begin{table*}
    \centering
    \caption{Column density ratios and excitation temperatures of the CN absorbers analysed in this study.}
    \label{tab:fitting_results}
    \begin{tabular}{lcccclcccc}
        \hline
                    & \multicolumn{4}{c}{Analysis based on $LSF_{\rm G}$} && \multicolumn{4}{c}{Analysis based on $LSF_{\rm V}$} \\
        \cline{2-5}\cline{7-10}
        CN Absorber & $\ronezero$ & $\sigma_{\rm lsf}$ & $\sigma_{\rm zl}$ & $T_{\rm 01}$ && $\ronezero$ & $\sigma_{\rm lsf}$ & $\sigma_{\rm zl}$ & $T_{\rm 01}$ \\
        \hline
        HD\,24534 & $0.4226\pm0.0168$ & $0.0014$ & $0.0027$ & $2.777\pm0.057$ && $0.4764\pm0.0150$ & $0.0007$ & $0.0027$ & $2.958\pm0.051$ \\
        HD\,27778 & $0.4073\pm0.0061$ & $0.0009$ & $0.0019$ & $2.726\pm0.022$ && $0.4392\pm0.0059$ & $0.0003$ & $0.0019$ & $2.832\pm0.021$ \\
        HD\,62542 ($^{12}{\rm C}^{14}{\rm N}$) & $0.3996\pm0.0156$ & $0.0139$ & $0.0270$ & $2.70\pm0.12$ && $0.7069\pm0.0075$ & $0.0039$ & $0.0270$ & $3.77\pm0.11$ \\
        HD\,62542 ($^{13}{\rm C}^{14}{\rm N}$) & $0.5083\pm0.0814$ & $0.0015$ & $0.0029$ & $2.94\pm0.27$ && $0.5302\pm0.0881$ & $0.0019$ & $0.0029$ & $3.01\pm0.29$ \\
        HD\,73882 & $0.3303\pm0.0172$ & $0.0056$ & $0.0108$ & $2.467\pm0.073$ && $0.5163\pm0.0097$ & $0.0013$ & $0.0108$ & $3.093\pm0.050$ \\
        HD\,147933 & $0.5086\pm0.0295$ & $0.0006$ & $0.0008$ & $3.07\pm0.10$ && $0.5256\pm0.0296$ & $0.0003$ & $0.0008$ & $3.13\pm0.10$ \\
        HD\,149757 & $0.4307\pm0.0094$ & $0.0007$ & $\cdots$ & $2.804\pm0.032$ && $0.4363\pm0.0095$ & $0.0001$ & $\cdots$ & $2.821\pm0.032$ \\
        HD\,152236 & $0.3872\pm0.0141$ & $0.0004$ & $0.0002$ & $2.659\pm0.047$ && $0.3962\pm0.0143$ & $0.0001$ & $0.0002$ & $2.689\pm0.048$ \\
        HD\,152270 & $0.70\pm0.13$ & $0.0005$ & $\cdots$ & $3.76^{+0.48}_{-0.45}$ && $0.70\pm0.13$ & $0.0001$ & $\cdots$ & $3.77^{+0.47}_{-0.45}$ \\
\hline
        \hline
                    & \multicolumn{4}{c}{Analysis based on $LSF_{\rm G}$} && \multicolumn{4}{c}{Analysis based on $LSF_{\rm V}$} \\
        \cline{2-5}\cline{7-10}
        CN Absorber & $\rtwoone$ & $\sigma_{\rm lsf}$ & $\sigma_{\rm zl}$ & $T_{\rm 12}$ && $\rtwoone$ & $\sigma_{\rm lsf}$ & $\sigma_{\rm zl}$ & $T_{\rm 12}$ \\
        \hline
        HD\,24534 & $0.0655\pm0.0268$ & $0.0002$ & $0.0001$ & $3.21^{+0.46}_{-0.73}$ && $0.0685\pm0.0276$ & $0.0001$ & $0.0001$ & $3.29^{+0.45}_{-0.71}$ \\
        HD\,27778 & $0.0389\pm0.0143$ & $0.0001$ & $0.0001$ & $2.81^{+0.29}_{-0.42}$ && $0.0402\pm0.0146$ & $0.0000$ & $0.0001$ & $2.84^{+0.28}_{-0.42}$ \\
        HD\,62542 & $0.0447\pm0.0033$ & $0.0003$ & $0.0007$ & $3.004\pm0.064$ && $0.0533\pm0.0038$ & $0.0002$ & $0.0007$ & $3.162\pm0.066$ \\
        HD\,73882 & $0.0405\pm0.0085$ & $0.0001$ & $0.0002$ & $2.91^{+0.16}_{-0.18}$ && $0.0445\pm0.0093$ & $0.0001$ & $0.0002$ & $2.99^{+0.17}_{-0.19}$ \\
        \hline
    \end{tabular}
\end{table*}

For a given absorption component, the column density of a given level is the same for all absorption lines. For example, the $R(1)$ and $P(1)$ lines both arise from the same lower level ($N''=1$), and therefore have the same total column density. We assume that the relative populations of the lower energy level of each transition are in thermal equilibrium. For example, transitions from the $N''=1, J''=3/2$ level are forced to be twice as strong as those from the $N''=1, J''=1/2$ level. Furthermore, unlike previous analyses, we perform a direct fit to the column density \emph{ratio} of the levels (cf. the left hand side of Equations~\ref{eqn:r1r0} and \ref{eqn:r2r1}). Some sightlines are well fit with a single absorption component, while others require multiple absorption components to obtain a satisfactory fit. All absorption components along a single sightline are assumed to be modelled with a single column density ratio; this is to avoid introducing covariance between the column density ratios of two blended components. After we finalise the model fitting, we perform an additional `validation fit', where each component has a separate column density ratio. This test allows us to check if all absorption components are consistent with the column density ratio inferred from the joint fit.

For sightlines with multiple cloud components, we force each cloud of the absorption model to have the same relative velocity. To account for wavelength calibration errors between each multiplet, we include an additional free model parameter in our analysis that allows for a velocity shift of each multiplet relative to the $R(0)$ barycentric velocity.

As mentioned in Section~\ref{sec:lsf}, an important part of our analysis is the treatment of the LSF. Most of the CN absorption lines are unresolved (or barely resolved) even with the high spectral resolution of the data reported here. When analysing the data, we perform two fits: one using $LSF(\lambda)_{\rm G}$, and another with $LSF(\lambda,\gamma=0)_{\rm V}$; as discussed in Section~\ref{sec:lsf}, we suggest that the `true' LSF profile lies somewhere in this range. To assess the impact of each LSF on the final column density ratios, we generated 20 LSFs based on the bootstrap fitting procedure described in Section~\ref{sec:lsf}. We then fit the data with each of these 20 line profiles and store the values of the column density ratios. This provides a distribution of column density ratios, where the width of this distribution provides an estimate of the systematic uncertainty of the column density ratio due to our modelling of each considered LSF.
To overcome initialisation bias, we initialise each parameter with a starting value randomly drawn from a uniform distribution over a specified interval. The intervals adopted in our analysis are:
(1) Doppler parameters in the range [0.3,0.8] km~s$^{-1}$;
(2) Column density is within a (linear) factor of three of the best-fit value; and
(3) Column density ratio, both $\ronezero$ and $\rtwoone$, in the range [0.2,0.6].
We repeat each fit 10 times (each with a different set of starting parameters) for every bootstrapped sample of both LSFs (i.e. a total of 400 fits per sightline). Based on these simulations, we found that the initialisation of parameters typically impacts the derived column density ratio at the level of $10^{-5}$; this systematic is considerably subdominant, and is therefore not considered further here.

Another possible cause of systematic uncertainty is the zero-level of the data, that could be due to either a poor background subtraction or a covering factor of the gas cloud that departs from unity. Stronger absorption lines are more affected by the zero-level of the data than weaker absorption lines. This is also true of the LSF. To assess the zero-level of the ESPRESSO data, we inspected the strong, saturated lines of Na\,\textsc{i}~$\lambda5889$\,\AA\ and confirm that these absorption lines are consistent with the zero-level to within the noise of the data ($<1\%$). We determine the zero-level of the core of the Na\,\textsc{i}~$\lambda5889$\,\AA\ absorption line, perform a fit to the CN absorption lines using this non-zero estimate of the zero-level, and calculate the difference between the column density ratio.\footnote{This intrinsically assumes that the zero-level is wavelength independent. Given the current data, we have no way to assess the validity of this assumption, but we note that the zero-level is always subdominant compared to other sources of uncertainty.} This difference is used as an estimate of the $1\sigma$ systematic uncertainty of the column density ratio $\ronezero$\ associated with the zero-level of the data. An estimate of the zero-level uncertainty ($\sigma_{\rm zl}$) and the LSF uncertainty ($\sigma_{\rm lsf}$) for each absorption line system is provided in Table~\ref{tab:fitting_results}.

Finally, we point out that our entire analysis is conducted blind, so that we do not reveal the final derived column density ratios of any system until the analysis of all systems is complete. Once we settled on the complete analysis strategy and finalised the fits, we unblind the column density ratios and do not perform any tweaks to the data reduction or analysis thereafter.\footnote{After we unblinded the results, we realised an error in the \textsc{alis} model files of HD\,27778, HD\,62542, and HD\,73882 that only affects the initialisation range of the $\rtwoone$ ratio. Specifically, our blinded runs initialised the $\rtwoone$ ratio in the range [0.2,0.6], while the intended range was [0.0,0.2]. After unblinding, we repeated the analysis with the intended initialisation range, and the results were unchanged. This further confirms that the initialisation of the parameters does not appear to significantly bias our results.}

\subsection{Individual Systems}

In the following subsections, we discuss the model fitting details of each system in turn. Each sightline is fit independently, so that we report a single excitation temperature per sightline. The LSFs that were used for each sightline are shown in the $R(0)$ panels of Figures~\ref{fig:hd24534}-\ref{fig:hd152270}, on the same scale as the data for comparison. All figures display the best-fitting model based on the $LSF(\lambda)_{\rm G}$ fits, since these fits were generally found to produce a lower reduced $\chi^{2}$ than the $LSF(\lambda,\gamma=0)_{\rm V}$ fits. In what follows, we only mention the statistical errors. A summary of the key fitting results, including estimates of the systematic uncertainties of each sightline, is provided in Table~\ref{tab:fitting_results}.

\subsubsection{HD\,24534}  
\begin{figure*}
	\includegraphics[width=\textwidth]{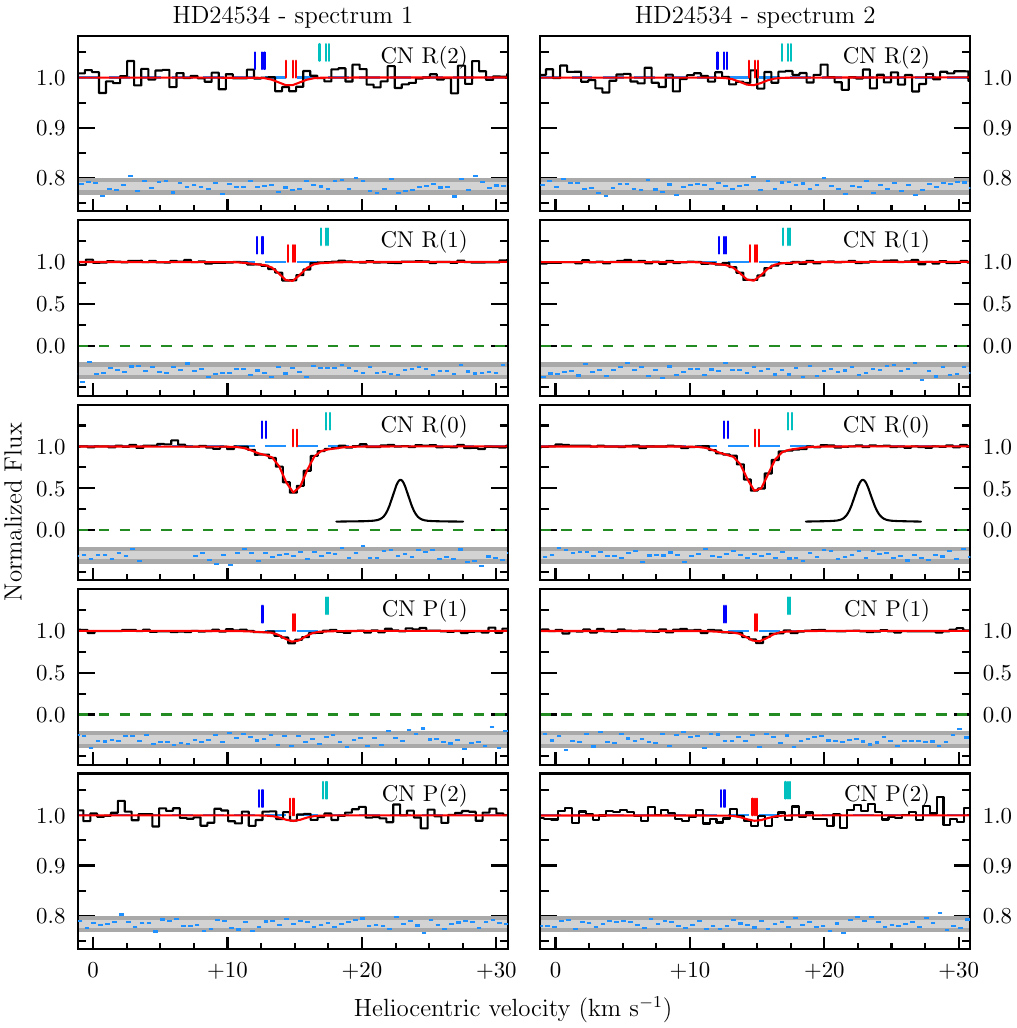}
    \caption{ESPRESSO data of CN absorption lines (black histogram) along the line-of-sight to HD\,24534. Overlaid in red is the best-fitting cloud model. The left and right set of panels show the data for the two ESPRESSO slices (see Section~\ref{sec:lsf} for details). Tick marks above the spectrum indicate the absorption components; each absorption component is represented by a different colour. Note that multiple tick marks of the same colour indicate the fine-structure of the absorption line. The continuum level and zero level are indicated with long blue dashed and short green dashed lines, respectively. The residuals, (data$-$model)/error, are shown at the bottom of each panel (blue markers), where dark and light shades of grey indicate $1\sigma$ and $2\sigma$ deviations about the model, respectively. The LSF that was used to model each spectrum is shown as a black curve in the middle panel on the same scale as the observations (i.e. the functional form of $LSF_{\rm G}$). Note the different y-axis scales that are used for the top and bottom panels to highlight the weak R(2) and P(2) lines.}
    \label{fig:hd24534}
\end{figure*}

The ESPRESSO data of HD\,24534 require a three component model, which consists of two satellite components at $v\simeq-2.3~{\rm km~s}^{-1}$ and $v\simeq+2.5~{\rm km~s}^{-1}$ relative to the strongest absorption component. Given that these satellite components are relatively weak and unresolved, we tie all three components to have the same Doppler parameter. We confidently detect the $R(0)$, $R(1)$, and $P(1)$ transitions, and possibly the $R(2)$ transitions in the strongest component at $2.5\sigma$. The best-fitting model is shown in Figure~\ref{fig:hd24534}, corresponding to a reduced chi-squared, $\chi^{2}/{\rm dof}=1.060$.

\begin{figure*}
	\includegraphics[width=\textwidth]{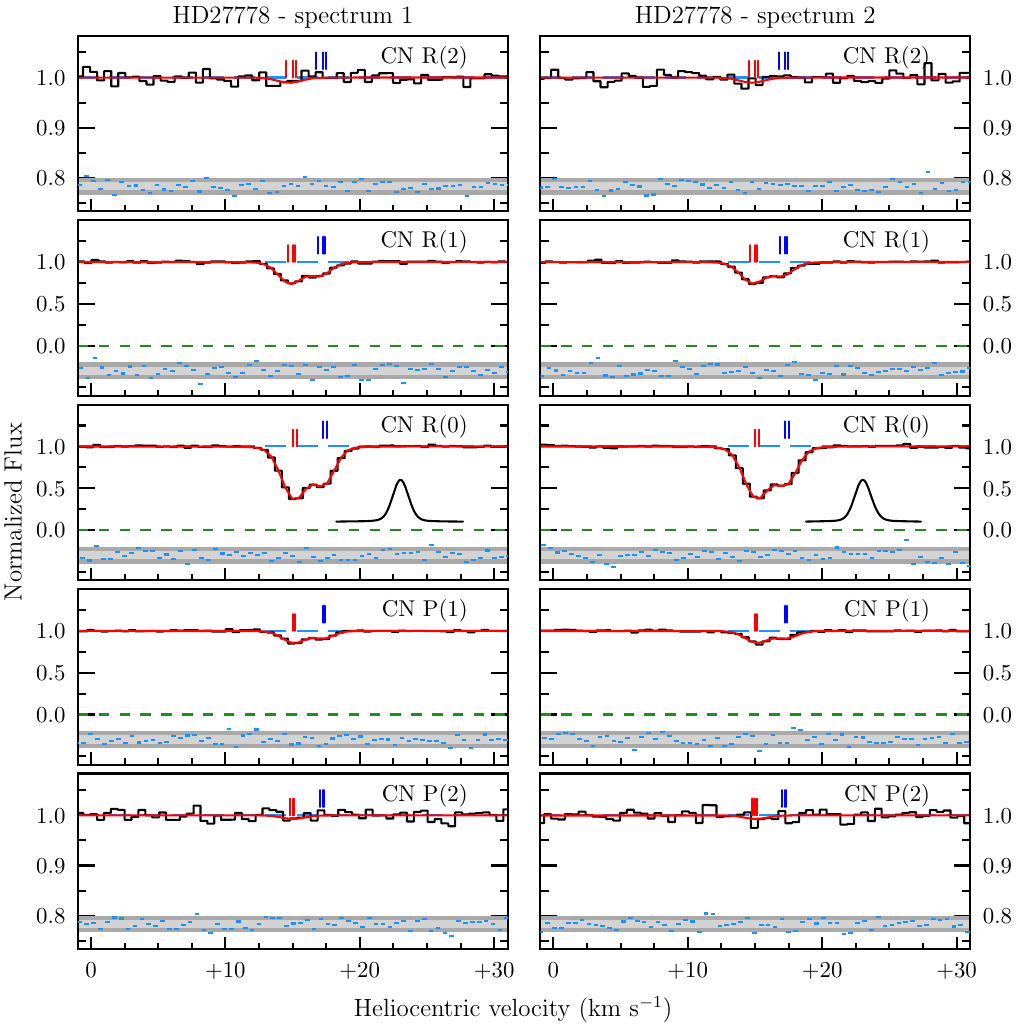}
    \caption{Same as Figure~\ref{fig:hd24534}, but for the absorption line system towards HD\,27778.}
    \label{fig:hd27778}
\end{figure*}

The total $N''=0$ column density of the strongest component is
$\log_{10}[N(0)/{\rm cm}^{-2}]=12.767\pm0.021$,
while the total Doppler parameter is $b=0.584\pm0.030~{\rm km~s}^{-1}$. A joint fit of the two ESPRESSO spectra, provides a column density ratio
$\ronezero=0.423\pm0.017$ using \lsfg. Analysing the two ESPRESSO spectra separately, we infer
$\ronezero=0.432\pm0.024$ and $\ronezero=0.413\pm0.024$, which are mutually consistent, and also consistent with the joint model fit.
For the \lsfv\ model, we obtain
$\ronezero=0.476\pm0.015$ for the joint fit of both spectra, and
$\ronezero=0.485\pm0.021$ and $\ronezero=0.468\pm0.021$ when analysing the
spectra separately. Overall, we find good agreement between the joint and separate fits, with the narrower \lsfv\ suggesting a column density ratio that is elevated by 0.053 ($\sim12$ percent) relative to \lsfg. This sightline also provides a measure of the $\rtwoone$ ratio, which is consistent between all analyses (see Table~\ref{tab:fitting_results}).

\subsubsection{HD\,27778}

The ESPRESSO data of HD\,27778 are well-characterised by a two component cloud model, with the components separated by $v\simeq2.2~{\rm km~s}^{-1}$. The data and best-fitting model (with $\chi^{2}/{\rm dof}=1.237$) are shown in Figure~\ref{fig:hd27778}. Based on the joint fit to both spectra and assuming that each component has the same $\ronezero$ value, the parameters of the strongest absorption components are
$\log_{10}[N(0)/{\rm cm}^{-2}]=12.8724\pm0.0089$ and $b=0.786\pm0.018~{\rm km~s}^{-1}$, while for the weaker component, we derive
$\log_{10}[N(0)/{\rm cm}^{-2}]=12.580\pm0.010$ and $b=0.567\pm0.030~{\rm km~s}^{-1}$.
Since there are two clear, well-defined absorption components, we can analyse the data in several different ways to test if each component and spectrum are consistent with the values we infer for the joint fit.

First consider the baseline measurement of the column density ratio, based on \lsfg, $\ronezero=0.4073\pm0.0061$. Analysing the two ESPRESSO spectra separately, we achieve consistent results, with $\ronezero=0.3996\pm0.0083$ and $\ronezero=0.4156\pm0.0089$. Alternatively, analysing the data jointly, but the components separately, we obtain $\ronezero=0.4081\pm0.0076$ and $\ronezero=0.405\pm0.013$, which are also mutually consistent.

Now considering the \lsfv\ results, the baseline for jointly fitting the data and the component column density ratios gives $\ronezero=0.4392\pm0.0059$. A separate analysis of the two spectra, but jointly fitting the component column density ratio yields $\ronezero=0.4311\pm0.0081$ and $\ronezero=0.4482\pm0.0086$. Meanwhile, a joint analysis of the data leads to the following column density ratios for the two components, $\ronezero=0.4416\pm0.0073$ and $\ronezero=0.433\pm0.013$, which are in good mutual agreement. Overall, regardless of how the data are analysed, we find consistent results for a given LSF, however the \lsfv\ results are elevated by $\sim8$ percent relative to the corresponding \lsfg\ values.

\subsubsection{HD\,62542}
\begin{figure*}
	\includegraphics[width=\textwidth]{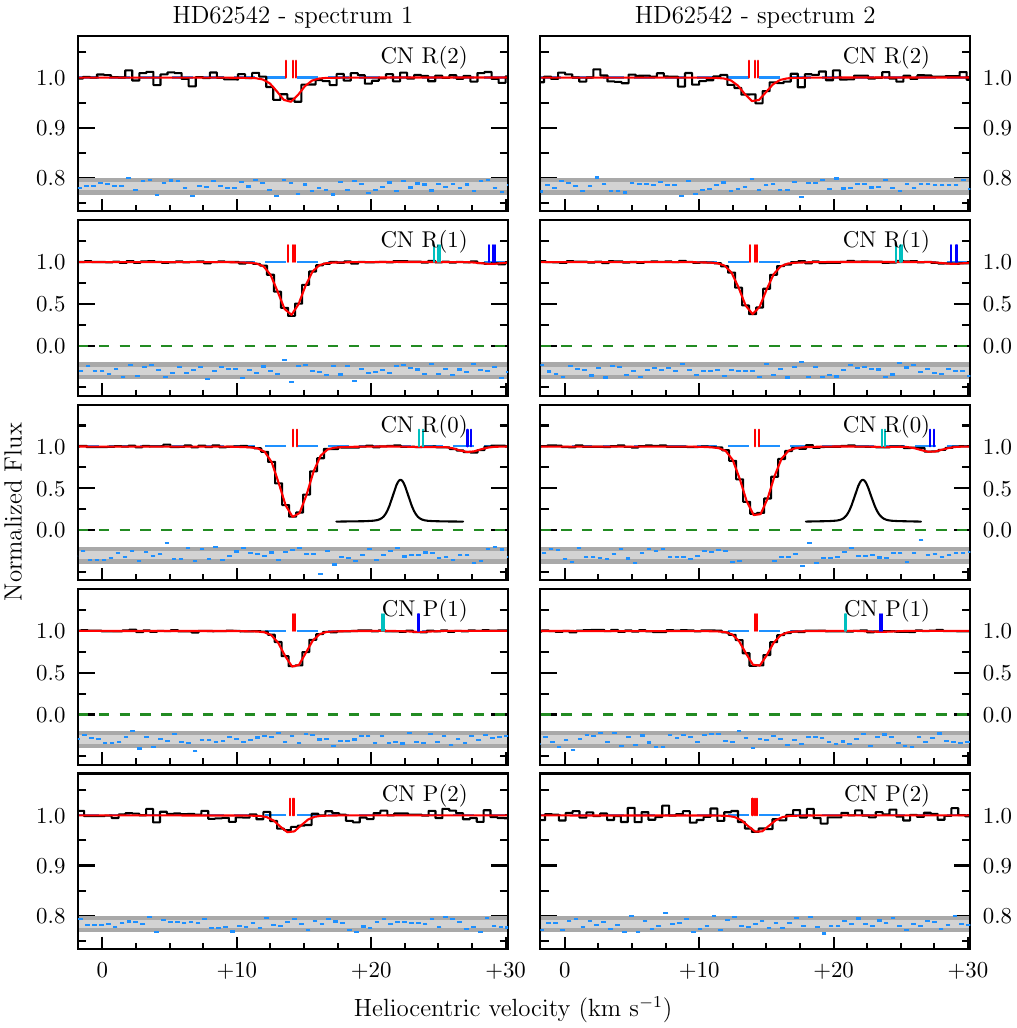}
    \caption{Same as Figure~\ref{fig:hd24534}, but for the absorption line system towards HD\,62542. The dark and light blue tick marks indicate isotopic absorption from $^{13}{\rm C}^{14}{\rm N}$ and $^{12}{\rm C}^{15}{\rm N}$, respectively (see Figure~\ref{fig:hd62542zoom} for a zoom-in of the isotope absorption).}
    \label{fig:hd62542}
\end{figure*}
\begin{figure*}
	\includegraphics[width=\textwidth]{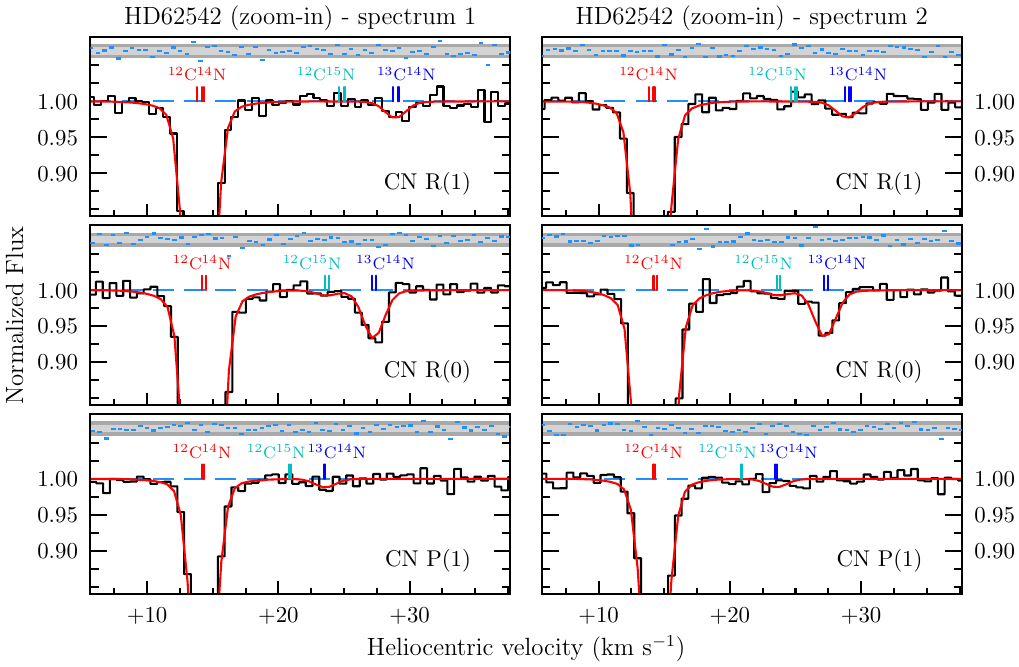}
    \caption{Same as Figure~\ref{fig:hd62542}, but showing a zoom-in of the CN isotope absorption features. Each isotope absorption line is identified with tick marks above each spectrum, as labelled.}
    \label{fig:hd62542zoom}
\end{figure*}

The CN absorber towards HD\,62542 consists of a single, strong component, where we detect the $R(0)$, $R(1)$, $R(2)$, $P(1)$, and $P(2)$ transitions. These data and the best-fitting model ($\chi^{2}/{\rm dof}=1.205$) are shown in Figure~\ref{fig:hd62542}. The $N(0)$ column density is $\log_{10}[N(0)/{\rm cm}^{-2}]=13.467\pm0.018$ and the Doppler parameter is $b=0.6417\pm0.0083~{\rm km~s}^{-1}$. Given the relatively large column density of this line, it is intrinsically saturated, and systematic uncertainties (particularly the uncertainty of the zero-level) dominate the total error budget (see Table~\ref{tab:fitting_results}).

We also find that the column density ratio is highly sensitive to the choice of LSF. The \lsfg\ analysis results in a column density ratio $\ronezero=0.400\pm0.016$, while the \lsfv\ analysis yields $\ronezero=0.7069\pm0.0075$ (i.e. elevated by $\sim75$ percent). Clearly, the LSF plays a significant role in the determination of the column density ratios of these strong, saturated lines. In this case, the FWHM of the two LSF values differ by just $\sim0.17~{\rm km~s}^{-1}$.

We also check if the results are consistent when we analyse each ESPRESSO spectrum separately. For this example, we find that one spectrum provides a good agreement with the joint fit ($0.401 \pm 0.019$) while the other spectrum exhibits a $1.7\sigma$ difference ($0.340 \pm 0.031$). This difference is perhaps not unexpected, because spectrum 2 suffers from a warm pixel near the core of the line profile that we have masked during the fitting procedure (see Figure~\ref{fig:hd62542}). As a result, spectrum 2 prefers a slightly higher $N(0)$ column density, corresponding to a lower $\ronezero$ ratio.

Along this sightline, we also detect absorption from some of the less abundant CN isotopes. Aside from the strong $^{12}{\rm C}^{14}{\rm N}$ absorption, we detect weak $^{13}{\rm C}^{14}{\rm N}$ and --- for the first time along this sightline --- a marginal detection of $^{12}{\rm C}^{15}{\rm N}$ absorption. We measure an isotope ratio of $^{12}{\rm C}{\rm N}/^{13}{\rm C}{\rm N}=55.4\pm8.5$, which is comparable to the weighted mean value ($^{12}{\rm C}{\rm N}/^{13}{\rm C}{\rm N}=67.5\pm1.0$) reported by \citet{Ritchey2011}. Currently, there are just four detections of $^{12}{\rm C}^{15}{\rm N}$ based on optical absorption lines, all of which are reported by \citet{Ritchey2015}. We report a marginal detection here, with an isotope ratio of ${\rm C}^{14}{\rm N}/{\rm C}^{15}{\rm N}=790\pm430$ ($1.8\sigma$ confidence), which is only weakly constrained, but still in agreement with the typical values reported by \citep[][${\rm C}^{14}{\rm N}/{\rm C}^{15}{\rm N}\sim300$]{Ritchey2015}. Further data of this sightline are needed before the detection of $^{12}{\rm C}^{15}{\rm N}$ can be confirmed. A zoom-in of the isotope absorption lines along this sightline is shown in Figure~\ref{fig:hd62542zoom}.


\subsubsection{HD\,73882}
\begin{figure*}
	\includegraphics[width=\textwidth]{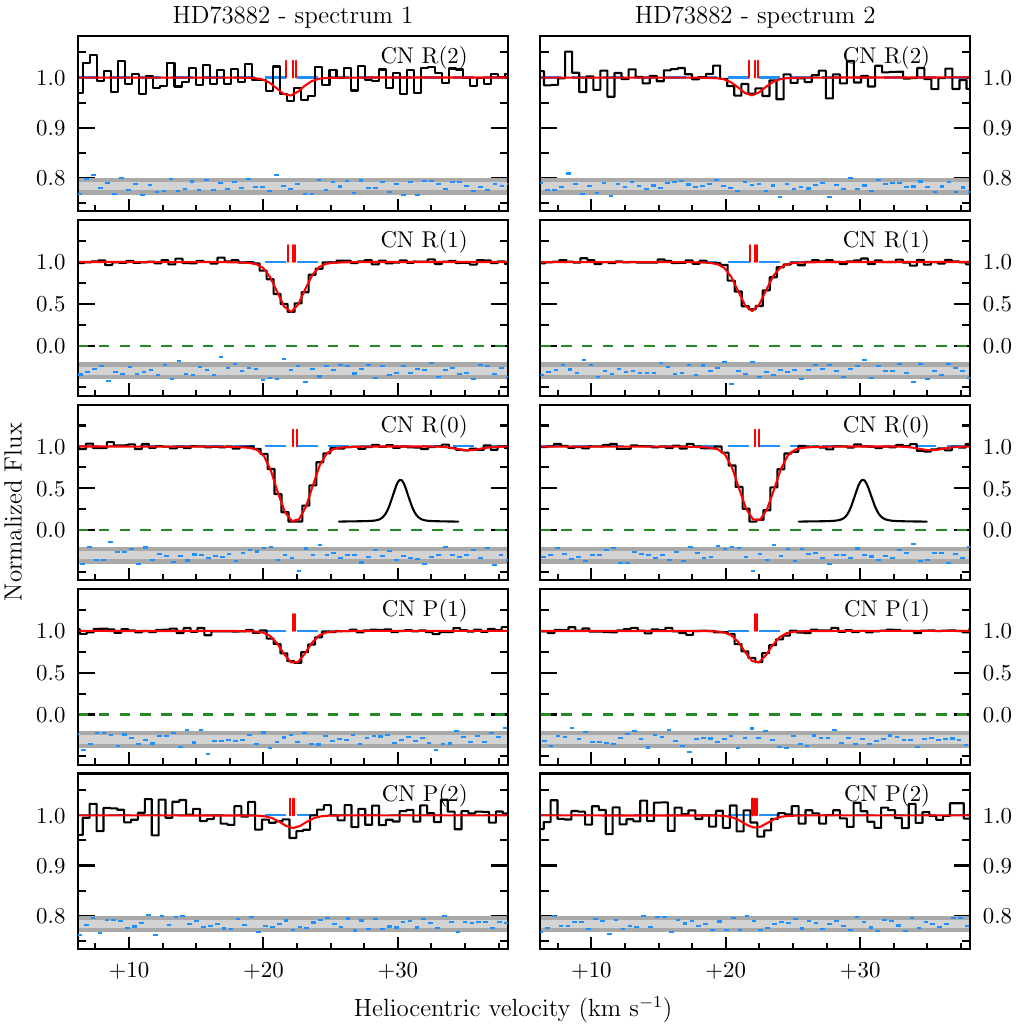}
    \caption{Same as Figure~\ref{fig:hd24534}, but for the absorption line system towards HD\,73882.}
    \label{fig:hd73882}
\end{figure*}

The ESPRESSO data of HD\,73882 reveals CN absorption lines from all transitions considered in this study, including $R(2)$, $R(1)$, $R(0)$, $P(1)$, and $P(2)$ (see Figure~\ref{fig:hd73882}). The best-fitting model consists of a single component, with an $N(0)$ column density $\log_{10}[N(0)/{\rm cm}^{-2}]=13.500\pm0.024$ and Doppler parameter $b=0.845\pm0.015~{\rm km~s}^{-1}$. This CN cloud is both the strongest and broadest absorber that we consider in this work. As a result, it is the most sensitive to the shape of the LSF, saturation, and improper accounting of the zero-level of the data (see Table~\ref{tab:fitting_results}). Based on the $\chi^{2}$ statistic, this model is also one of the poorest fits considered in this paper ($\chi^{2}/{\rm dof}=1.494$). A joint fit of the data yields a column density ratio $\ronezero=0.330\pm0.017$, which is the lowest value reported here for the \lsfg\ analysis. The \lsfv\ results are considerably higher, with $\ronezero=0.5163\pm0.0097$, indicating the sensitivity of this result to the shape of the LSF. Analysing the two ESPRESSO data separately (assuming the \lsfg\ analysis), yields consistent results $\ronezero=0.334 \pm 0.023$ and $\ronezero=0.326 \pm 0.026$. This ratio is significantly below the ratio that would be expected for level populations that are in equilibrium with CMB photons.

After we unblinded the fitting results, we realised that the CN \axband\ (2,0) band is covered by the ESPRESSO data of this sightline. This band consists of weaker CN absorption lines that are less likely to suffer from biases due to line saturation. We analysed these lines using the same approach as used for the \band\ (0,0) band data. An independent fit to the \axband\ (2,0) lines revealed a narrowed Doppler parameter ($b=0.71\pm0.11$). The $N(1)$ column density was found to be almost identical to the \band\ (0,0) band data, while the $N(0)$ column density was $\log_{10}[N(0)/{\rm cm}^{-2}]=13.403\pm0.076$, leading to a ratio $\ronezero=0.416 \pm 0.052$; this agrees very well with the results of \citet{Ritchey2011}. Alternatively, a joint fit to all available data yields an acceptable fit ($\chi^{2}/{\rm dof}=1.258$) that is consistent with the \band\ (0,0) band only results, indicating that the parameters are driven by the \band\ (0,0) band data. We therefore conclude that the $N(0)$ column density of this system appears to be affected by saturation of the \band\ (0,0) $R(0)$ line. This system is discussed further in Section~\ref{sec:discussion}.


\subsubsection{HD\,147933~=~$\rho$ Oph A}
\begin{figure*}
	\includegraphics[width=\textwidth]{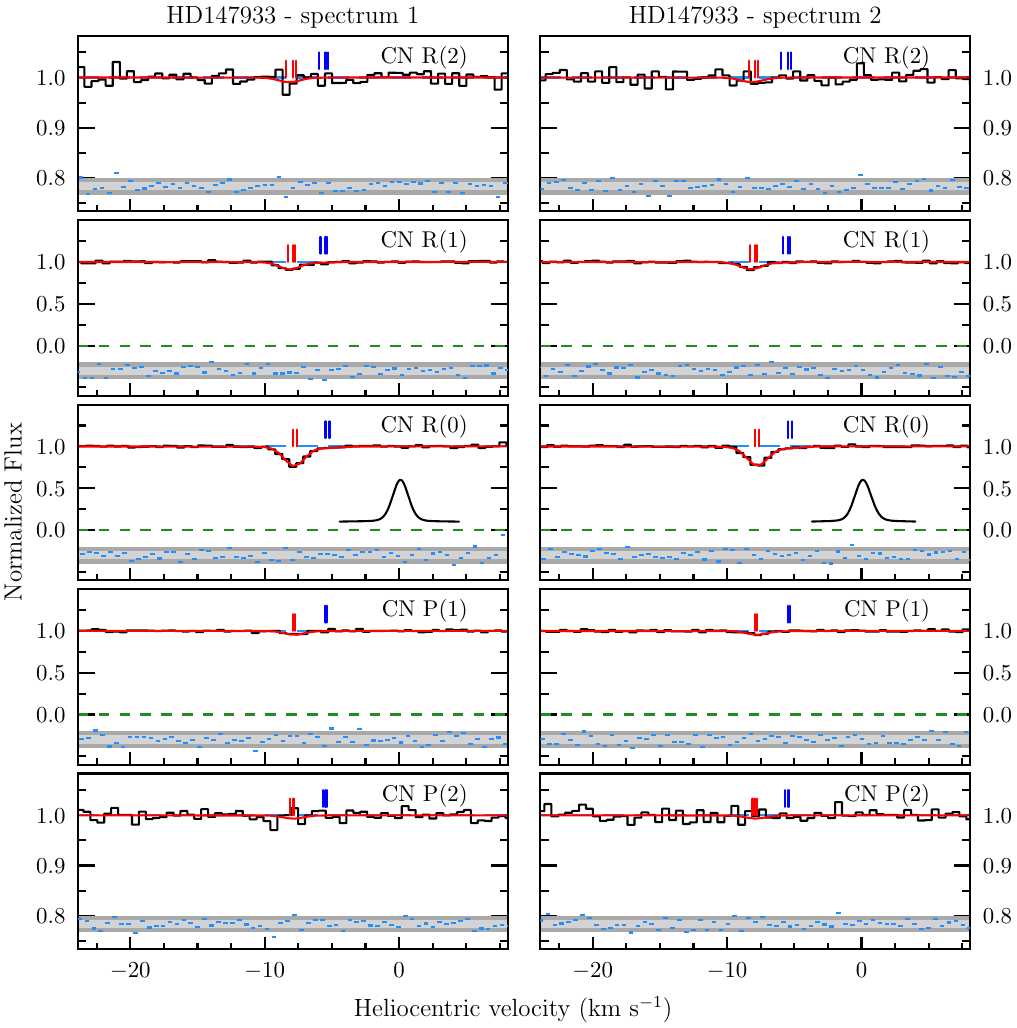}
    \caption{Same as Figure~\ref{fig:hd24534}, but for the absorption line system towards HD\,147933.}
    \label{fig:hd147933}
\end{figure*}

The CN absorber towards HD\,147933 is a relatively weak system, and requires a two component absorption model to obtain an adequate fit ($\chi^{2}/{\rm dof}=1.207$; see Figure~\ref{fig:hd147933}). The primary component has an $N(0)$ column density $\log_{10}[N(0)/{\rm cm}^{-2}]=12.219\pm0.035$ and Doppler parameter $b=0.560\pm0.051~{\rm km~s}^{-1}$. The satellite absorption component is located at $\sim+2.5~{\rm km~s}^{-1}$ relative to the primary component, and is weaker by a factor of $\sim20$. Given the low column density of the satellite component, we decided to tie the corresponding Doppler parameter to that of the primary component. A joint fit to the ESPRESSO data yields a column density ratio $\ronezero=0.509\pm0.030$ for the \lsfg\ analysis (note the \lsfv\ results are within $1\sigma$ of this result). We have also analysed the ESPRESSO spectra separately, and find that the results are mutually consistent, with values $\ronezero=0.539\pm0.041$ and $\ronezero=0.474\pm0.042$. Finally, we note that the $R(2)$ line is marginally detected ($\sim1.9\sigma$ confidence), so we do not report a $T_{12}$ value of this absorber.


\subsubsection{HD\,149757~=~$\zeta$ Oph}
\begin{figure}
	\includegraphics[width=\columnwidth]{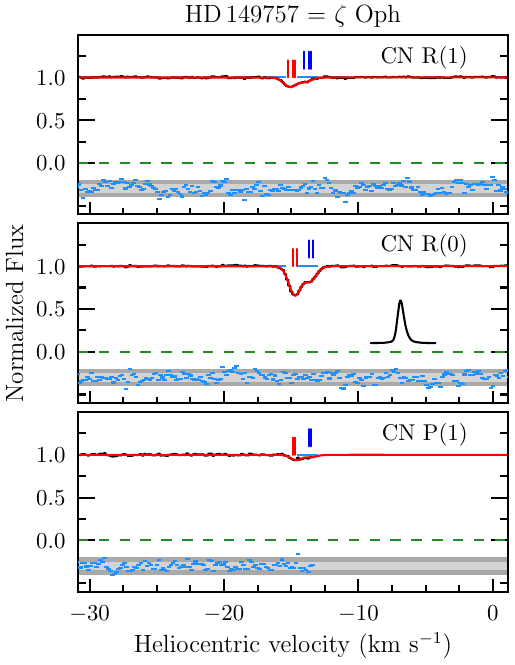}
    \caption{Same as Figure~\ref{fig:hd24534}, but for the absorption line system towards HD\,149757 ($=\zeta$ Oph), and based on AAT/UHRF data.}
    \label{fig:hd149757}
\end{figure}

The archival UHRF observations of HD\,149757 are perhaps the most valuable in terms of both the S/N and the spectral resolution. These data were originally analysed by \citet{Crawford1994}, and like those authors, we fit the CN absorption system with a two component model. The $P(1)$ absorption line falls at the edge of the detector; part of the line profile is not recorded, but we nevertheless fit the available data. We are able to find a consistent fit to all detected absorption lines, with $\chi^{2}/{\rm dof}=0.989$. The two absorption components are separated by just $\sim1.2~{\rm km~s}^{-1}$, with column densities $\log_{10}[N(0)/{\rm cm}^{-2}]=12.163\pm0.014$ and $\log_{10}[N(0)/{\rm cm}^{-2}]=11.726\pm0.021$, and corresponding Doppler parameters $b=0.440\pm0.013~{\rm km~s}^{-1}$ and $b=0.356\pm0.030~{\rm km~s}^{-1}$. Note that the LSF shown in the middle panel of Figure~\ref{fig:hd149757} is substantially asymmetric.

Our baseline fit to this system assumes that both components have an identical $\ronezero$ column density ratio, with a best-fitting value $\ronezero=0.4307\pm0.0094$ for the \lsfg\ analysis. We also note that this value is relatively insensitive to the adopted LSF, for example, the \lsfv\ analysis yields $\ronezero=0.4363\pm0.0095$. We have also performed a fit that allows both components to have an independent $\ronezero$ ratio. In this case, the leftmost component has a value $\ronezero=0.435\pm0.011$, while the rightmost component has a value $\ronezero=0.407\pm0.028$ --- both values are in good mutual agreement (i.e. within $\sim1\sigma$) and also agree with the joint fit. Finally, we note that we do not have an estimate of the zero-level uncertainty for this system, but given that the absorption lines are very weak, the zero-level is not expected to significantly affect the result.


\subsubsection{HD\,152236~=~$\zeta^{1}$ Sco}
\begin{figure*}
	\includegraphics[width=\textwidth]{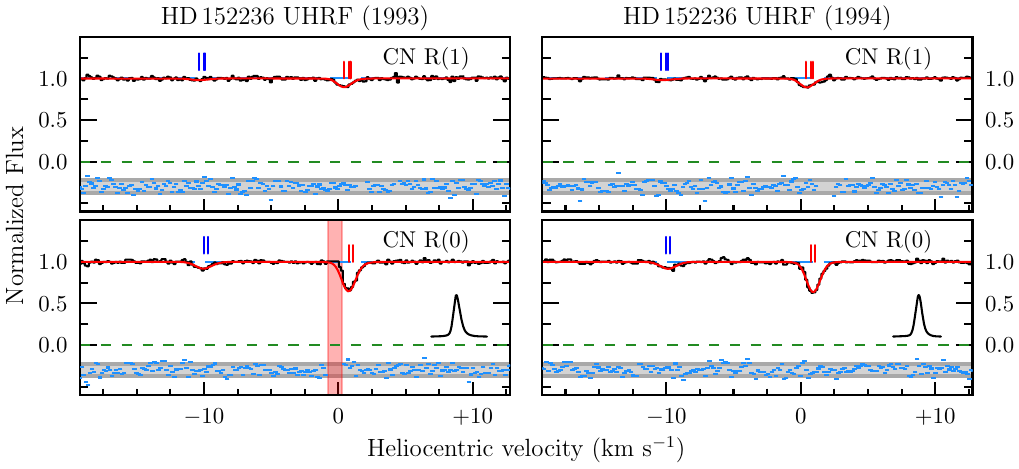}
    \caption{Same as Figure~\ref{fig:hd24534}, but for the absorption line system towards HD\,152236 ($=\zeta^{1}$ Sco), and based on AAT/UHRF data. See also, Figure~\ref{fig:hd152236esp}. Note that the vertical red band indicates a series of pixels that are impacted by a cosmic ray, and were masked during the fitting procedure.}
    \label{fig:hd152236uhrf}
\end{figure*}

\begin{figure*}
	\includegraphics[width=\textwidth]{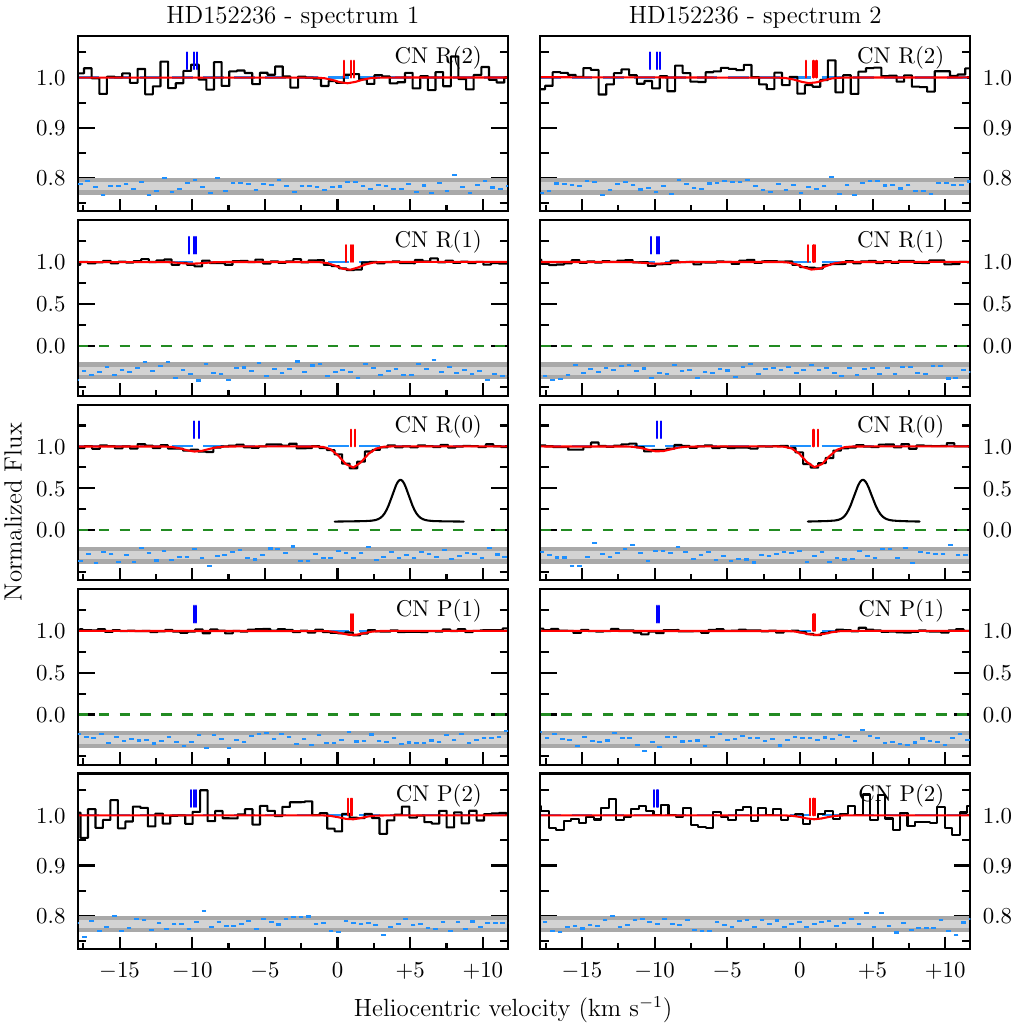}
    \caption{Same as Figure~\ref{fig:hd24534}, but for the absorption line system towards HD\,152236 ($=\zeta^{1}$ Sco). See also, Figure~\ref{fig:hd152236uhrf}.}
    \label{fig:hd152236esp}
\end{figure*}

The sightline towards HD\,152236 is unique among our sample, since it has been observed with both UHRF (Figure~\ref{fig:hd152236uhrf}) and ESPRESSO (Figure~\ref{fig:hd152236esp}). The line-of-sight intersects two weak CN absorption components separated by $10.8~{\rm km~s}^{-1}$. The stronger component has an $N(0)$ column density $\log_{10}[N(N(0))/{\rm cm}^{-2}]=12.222\pm0.022$ and Doppler parameter $b=0.52\pm0.02$, and is significantly detected in the $R(0)$, $R(1)$, and $P(1)$ transitions. The weaker (blue-shifted) component has an $N(0)$ column density $\log_{10}[N(0)/{\rm cm}^{-2}]=11.581\pm0.033$ and Doppler parameter $b=0.69\pm0.07~{\rm km~s}^{-1}$. The weaker component is well-detected in the $R(0)$ line, but only weakly detected in the $R(1)$ and $P(1)$ lines. Our baseline best-fitting model ($\chi^{2}/{\rm dof}=1.232$) for this sightline is based on a single joint fit to all available data, and assumes both components have the same $\ronezero$ ratio. Our best-fitting $\ronezero$ column density ratio, based on the \lsfg\ analysis, is $\ronezero=0.387\pm0.014$. A consistent result is obtained for the \lsfv\ analysis, with $\ronezero=0.396\pm0.014$.

We have a total of four datasets of this sightline that we can use to test for consistency; these data consist of two UHRF observing runs (during 1993 and 1994), and two ESPRESSO spectra (one from each slice that was acquired simultaneously). We note that the 1993 UHRF data are affected by a cosmic ray event near the strongest CN $R(0)$ absorption component; the affected data were masked during the fitting procedure. In what follows, we only compare the \lsfg\ results. We report a good mutual agreement between the 1993 and 1994 UHRF data, with $\ronezero=0.392\pm0.050$ and $\ronezero=0.373\pm0.018$, respectively. These values also agree with our baseline model that jointly fits all available data (recall, $\ronezero=0.387\pm0.014$).

The ESPRESSO data, however, are somewhat inconsistent with each other (at the $\sim2\sigma$ level), with values $\ronezero=0.495\pm0.068$ and $\ronezero=0.326\pm0.055$, and only marginally consistent with the joint fit. This difference can be explained by the ESPRESSO modelling; the Doppler parameters of the two components tend towards either unfeasibly low values ($\sim0.01~{\rm km~s}^{-1}$) or relatively high values ($\sim1~{\rm km~s}^{-1}$). We suspect this is due to the poorer line detection significance of the ESPRESSO data, since the lines are unresolved and the data are of relatively low S/N, relative to the UHRF data. As a result, the noise is impacting the shape of the (unresolved) line profile, and forcing the Doppler parameters into unrealistic parameter space. Fortunately, a joint fit to just the ESPRESSO data provides a consistent value, $\ronezero=0.416\pm0.043$.
Given that the UHRF data agree with the baseline result, we suspect that the UHRF data strongly constrain the cloud model of the joint fit, and the ESPRESSO data are primarily helping to reduce the statistical error. We note that neither the zero-level nor the LSF choice contributes significantly to the total error budget.


\subsubsection{HD\,152270}
\begin{figure}
	\includegraphics[width=\columnwidth]{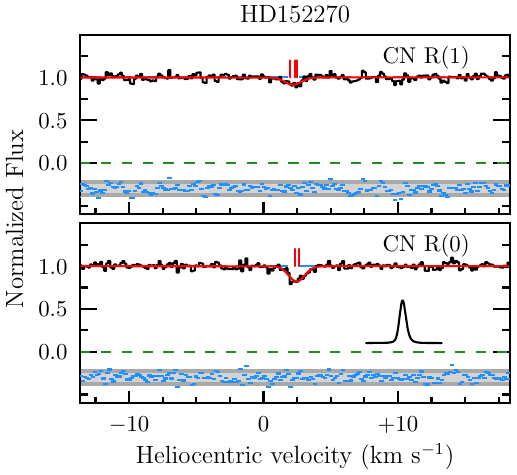}
    \caption{Same as Figure~\ref{fig:hd24534}, but for the absorption line system towards HD\,152270.}
    \label{fig:hd152270}
\end{figure}

The UHRF data of HD\,152270 are the highest spectral resolution, but also the lowest S/N, of all spectra considered in this paper. These data only cover the $R(0)$ and $R(1)$ lines, and the absorption appears to consist of a single weak absorption component (see Figure~\ref{fig:hd152270}). The best-fitting model ($\chi^{2}/{\rm dof}=0.813$) favours a weak absorber with an $N(0)$ column density $\log_{10}[N(0)/{\rm cm}^{-2}]=11.94\pm0.11$ and Doppler parameter $b=0.67\pm0.09~{\rm km~s}^{-1}$. The derived $\ronezero$ column density ratio for the \lsfg\ analysis is $\ronezero=0.70\pm0.13$, which is identical to the result derived with the \lsfv\ analysis. The statistical errors dominate the error budget of the $\ronezero$ value. We also note that we are unable to estimate the zero-level systematic uncertainty of these data, but we do not expect this to be significant since the absorption lines are very weak.

\subsection{Excitation temperatures}
\label{sec:tex}

A summary of the best-fitting column density ratios of each sightline, and for both LSF analyses, are collected in Table~\ref{tab:fitting_results}. To calculate the posterior distributions of the excitation temperature based on these column density ratios, we perform a Markov chain Monte Carlo (MCMC) analysis using the \textsc{emcee} software \citep{Foreman-Mackey2013}. We adopt a uniform prior on the excitation temperature in the range $0\leq T_{01}/{\rm K}\leq5$. We randomly initialise 50 walkers within this interval, and run the MCMC for 5000 steps, with a conservative burn-in of 500 steps, and thinned by a factor of 15. We confirmed that this set of parameters produced converged chains by computing the auto-correlation time. The inferred median, and $1\sigma$ confidence limits on the excitation temperatures are provided in Table~\ref{tab:fitting_results}.

\begin{figure*}
	\includegraphics[width=\columnwidth]{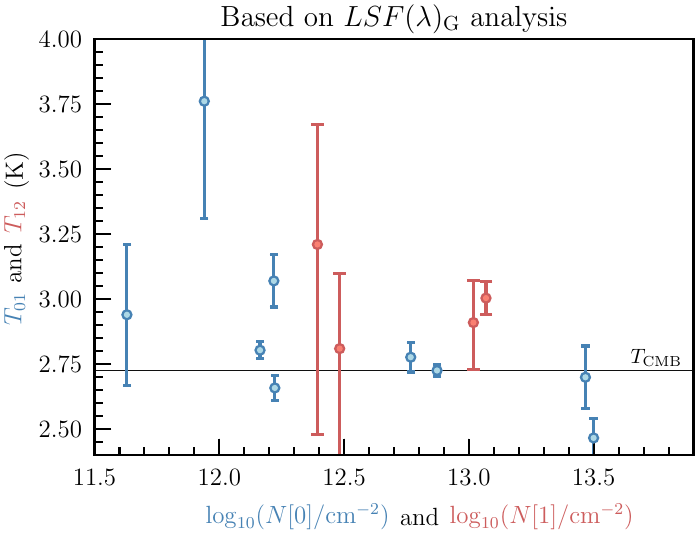}
        \includegraphics[width=\columnwidth]{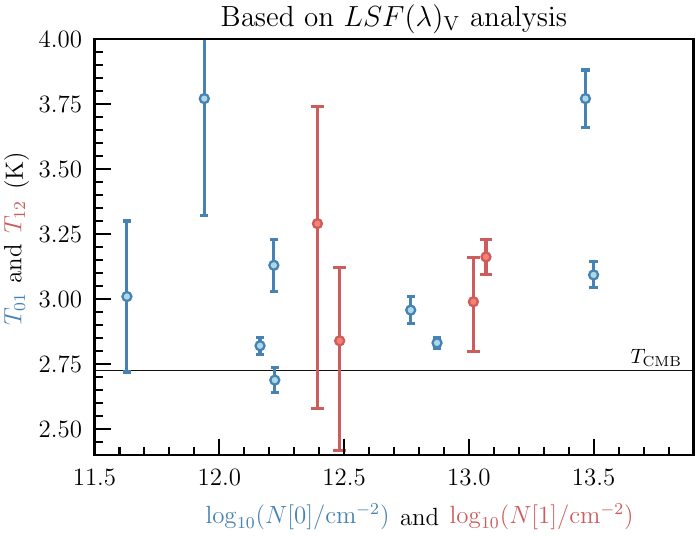}
    \caption{The excitation temperature $T_{01}$ (blue symbols) and $T_{12}$ (red symbols) as a function of the column density of the lower level. The black line is the direct measurement of the CMB temperature, $T_{\rm CMB,0}=2.7260 \pm 0.0013~{\rm K}$ \citep{Fixsen2009}. The left and right panels show the analysis results based on the \lsfg\ and \lsfv\ models of the instrument resolution, respectively.}
    \label{fig:tcmb}
\end{figure*}

\subsection{Comparison with previous work}
\label{sec:comparison}

Many of the sightlines considered in this paper were also analysed by \citet{Ritchey2011}, using data of generally higher S/N and lower spectral resolution. By comparing the results of the sightlines that are in common between these two studies, we can try to understand how human choices and differences in the data and analysis strategies can affect the final derived results. A summary of these results are collected in Table~\ref{tab:comparison}, where we list the \lsfg\ results since these offer the closest comparison to the \citet{Ritchey2011} analysis. Most of the measurements exhibit only minor differences at the $1-2\sigma$ level, while one of the measurements (HD\,147933) exhibits a $\sim4\sigma$ difference. Given that the Doppler parameters rarely agree within $1\sigma$, this disagreement propagates to larger differences in the column densities of the strong (and often saturated) CN lines, particularly the $R(0)$ line. Clearly, to further understand the physical conditions of the CN-bearing gas, we must pin down the instrumental broadening, and Doppler parameters of CN absorbers. This should be considered an important goal with the (known) Milky Way absorbers, before attempting to conduct this experiment with extragalactic and high redshift absorbers, where ultra-high spectral resolution will not be feasible.

\begin{table*}
    \centering
    \caption{CN absorber comparison between our results and \citet{Ritchey2011}}
    \label{tab:comparison}
    \begin{tabular}{lccclccc}
        \hline
                    & \multicolumn{3}{c}{Our Work} && \multicolumn{3}{c}{\citet{Ritchey2011}} \\
        \cline{2-4}\cline{6-8}
        CN Absorber & $N[0]$ & $b$ & $T_{\rm 01}$ && $N[0]$ & $b$ & $T_{\rm 01}$  \\
                    & $(10^{12}~{\rm cm}^{-2})$ & ${\rm km~s}^{-1}$ & $({\rm K})$ && $(10^{12}~{\rm cm}^{-2})$ & ${\rm km~s}^{-1}$ & $({\rm K})$  \\
        \hline
        HD\,73882 & $31.7\pm1.8$ & $0.845\pm0.015$ & $2.467\pm0.073$ && $27.66\pm0.04$ & $0.88$ & $2.631\pm0.004$ \\
        HD\,147933 & $1.66\pm0.13$ & $0.560\pm0.051$ & $3.07\pm0.10$ && $1.77\pm0.01$ & $0.42$ & $2.657\pm0.026$ \\
        HD\,149757 & $1.456\pm0.047$ & $0.440\pm0.013$ & $2.804\pm0.032$ && $1.61\pm0.02$ & $0.60$ & $2.702\pm0.042$ \\
        HD\,149757 & $0.532\pm0.026$ & $0.356\pm0.030$ & $2.804\pm0.032$ && $0.45\pm0.02$ & $0.40$ & $2.702\pm0.042$ \\
        HD\,152236 & $1.669\pm0.085$ & $0.52\pm0.02$ & $2.659\pm0.047$ && $2.21\pm0.02$ & $0.40$ & $2.588\pm0.033$ \\
        HD\,152236 & $0.382\pm0.029$ & $0.69\pm0.07$ & $2.659\pm0.047$ && $0.38\pm0.01$ & $0.54$ & $2.454\pm0.178$ \\
        \hline
    \end{tabular}
\end{table*}

\section{Discussion}
\label{sec:discussion}

All excitation temperature measurements are shown in Figure~\ref{fig:tcmb} as a function of the lower level column density. It is worth noting that when the column density of the lower level is $N\lesssim10^{12.5}~{\rm cm}^{-2}$, the difference between the two LSF analyses is minimal; the difference between the two LSF analyses increases for increasingly strong lines. This suggests that the most reliable excitation temperatures are to be derived from the lower range of CN column densities. Nevertheless, even the weak lines appear to exhibit an excess `intrinsic' dispersion beyond the quoted error budget. To quantify the level of intrinsic scatter, we construct a model to extract a measure of the unaccounted for systematic uncertainty.

Suppose that each column density ratio that we measure, $r_{i}\pm\sigma_{i}$, has a true value $r_{t}$. The probability that our measurement arises from this true value is
\begin{equation}
    {\rm Pr}(r_{i}|r_{t}) = \frac{1}{\sqrt{2\pi}\sigma_{i}}\exp\bigg(-\frac{(r_{i}-r_{t})^{2}}{2\sigma_{i}^{2}}\bigg)
\end{equation}
where the true value is assumed to be drawn from a Gaussian distribution with a central value $R$ and an intrinsic (i.e. ``systematic'') scatter $\sigma_{\rm sys}$. In this case, the probability that a given true value ($r_{t}$) is drawn from this distribution is
\begin{equation}
    {\rm Pr}(r_{t}|R) = \frac{1}{\sqrt{2\pi}\sigma_{\rm sys}}\exp\bigg(-\frac{(r_{t}-R)^{2}}{2\sigma_{\rm sys}^{2}}\bigg)
\end{equation}
By integrating over all possible true values, we arrive at the probability of obtaining a measured column density ratio, $r_{i}$, given a Gaussian distribution of values with an intrinsic scatter:
\begin{equation}
    {\rm Pr}(r_{i}|R) = \frac{1}{\sqrt{2\pi(\sigma_{i}^{2}+\sigma_{\rm sys}^{2})}}\exp\bigg(-\frac{(r_{i}-R)^{2}}{2(\sigma_{i}^{2}+\sigma_{\rm sys}^{2})}\bigg)
\end{equation}
and the corresponding log-likelihood function is of the form
\begin{equation}
    \label{eqn:likelihood}
    {\cal L} = \log\bigg[\prod_{i}~{\rm Pr}(r_{i}|R)]\bigg]
\end{equation}
where the true value, $R$, is given by the LHS of Equation~\ref{eqn:r1r0}. Our model therefore contains two free parameters: the excess uncertainty ($\sigma_{\rm sys}$) and the excitation temperature (either $T_{01}$ or $T_{12}$). Using the above log-likelihood function, we used \textsc{emcee} with the same configuration as described in Section~\ref{sec:tex} to calculate the posterior distributions of the typical excitation temperature and the associated intrinsic scatter associated with the measured column density ratios (for the \lsfg\ analysis):
\begin{equation}
    \label{eqn:t01result}
    T_{01} = 2.769^{+0.084}_{-0.072}~{\rm K}\\
    \sigma_{\rm sys} = 0.056^{+0.031}_{-0.019}
\end{equation}
Note that $\sigma_{\rm sys}$ is the intrinsic scatter associated with the measured column density ratios and not an intrinsic scatter associated with the excitation temperature. $\sigma_{\rm sys}$ therefore represents a $\sim10$ percent uncertainty of the column density ratio due to systematics that are presently unaccounted for. We also note that a simple weighted mean of the $T_{01}$ values listed in Table~\ref{tab:fitting_results} (i.e. set $\sigma_{\rm sys}=0$) gives $\langle T_{01}\rangle = 2.739\pm0.015~{\rm K}$; this uncertainty is a factor of $\sim6$ more precise than the value of $T_{01}$ that accounts for $\sigma_{\rm sys}$. Given that $\sigma_{\rm sys}$ is significantly non-zero, we stress the importance of the intrinsic scatter term when calculating the most likely mean value, instead of only quoting a weighted mean value.

Our determination of $T_{01}$ represents a reliable $\sim3$ percent determination of the typical CN excitation temperature of diffuse molecular clouds in the Milky Way. Our measure agrees with the direct measurement of the CMB temperature, $T_{\rm CMB,0}=2.7260\pm0.0013~{\rm K}$, by \citet{Fixsen2009}. Our measurements suggest that local sources of excitation do not significantly alter the CN level populations in our sample of absorbers, relative to the CMB. However, the source of intrinsic scatter is not currently known; to fully understand the intrinsic scatter a larger sample of systems may be required. The two most likely causes include sightline dependent excitation, or issues with the LSF and cloud modelling. To test these two possibilities, we can compare our result to those of other analyses reported in the literature. If $\sigma_{\rm sys}$ is identical between any two studies, it is more likely to be intrinsic to the CN absorbers, since different authors analyse their samples in different ways. This leads to modelling choices that are more likely to lead to a different set of systematic errors.

For this comparison, we consider the most recent CN sample reported by \citep{Ritchey2011}, who report a weighted mean value $\langle T_{01}\rangle = 2.754\pm0.002$. We note that a significant fraction ($\sim62$ percent) of their sample has an $R(0)$ column density that is $\lesssim10^{12.5}~{\rm cm}^{-2}$, which is the regime we find very little change to the values between the \lsfg\ and \lsfv\ analyses. Furthermore, their data are of generally higher S/N ratio, albeit at somewhat lower spectral resolution. If we apply our likelihood model (Equation~\ref{eqn:likelihood}) to their data, we find values:
\begin{equation}
    \label{eqn:ritchey}
    T_{01} = 2.717^{+0.036}_{-0.039}~{\rm K}\\
    \sigma_{\rm sys} = 0.038^{+0.008}_{-0.012}
\end{equation}
By simultaneously modelling the intrinsic scatter, we find a typical excitation temperature that is more consistent with the CMB value (compared to the weighted mean). Furthermore, the intrinsic dispersion is significantly non-zero, and is somewhat lower than the value that we derive for our sample. We also note that both studies are consistent within $1\sigma$. We therefore conclude that both our study and \citet{Ritchey2011} appear to suffer from an unknown systematic effect, despite the care that has been taken in both analyses. If this systematic uncertainty is due entirely to local sources of excitation, then we conclude $\Delta T_{\rm loc}\leq0.054\,{\rm K}~(2\sigma)$.

Repeating our likelihood calculation with the results of the \lsfv\ analysis, we find
\begin{equation}
    T_{01} = 3.06\pm0.13~{\rm K}\\
    \sigma_{\rm sys} = 0.107^{+0.037}_{-0.026}
\end{equation}
which is significantly ($\sim10$ percent) elevated relative to the \lsfg\ analysis. Given the agreement between our \lsfg\ analysis and the independent work by \citet{Ritchey2011}, we propose that our \lsfg\ analysis is more reliable than our \lsfv\ analysis. We also note that the \lsfg\ analysis is the `standard' approach; the instrument resolution is usually assumed to be accurately reflected by the widths of unresolved ThAr lines. The \lsfv\ analysis nevertheless illustrates the sensitivity of this measurement to small changes in the LSF. Finally, given that the \lsfg\ and \lsfv\ analyses represent the `extremes' of the LSF, we can place a confident ($2\sigma$) limit on the typical excitation temperature of Milky Way sightlines, $2.62 \leq T_{01}/{\rm K} \leq 3.32$.

\subsection{Simultaneous Joint Fit}

As a final step, we perform a single, joint fit to all eight sightlines simultaneously using the initial model setup described in Section~\ref{sec:profilefitting}. The only change to the blinded models described above is that all absorbers and components are assumed to have a single $\ronezero$ and a single $\rtwoone$ value. We also include a separate model parameter to calculate the column density ratio of $^{13}{\rm C}^{14}{\rm N}$. The optimised model parameters provide a good fit to the data ($\chi^{2}/{\rm dof}=1.186$) and, as previously, we find that the initialisation bias does not impact the results. The best-fitting value of the column density ratios are, $\ronezero=0.4074\pm0.0042$ and $\rtwoone=0.0445\pm0.0030$, which correspond to the following excitation temperatures:
\begin{eqnarray}
    T_{01} = 2.725\pm0.015~{\rm K}\\
    T_{12} = 3.002\pm0.055~{\rm K}
\end{eqnarray}
The derived $T_{01}$ is a 0.55\%\ measure --- this value is a factor of $\sim5$ more precise than the value reported by analysing each system separately (cf. Equation~\ref{eqn:t01result}). Furthermore, our joint fit is more consistent with both \citet{Ritchey2011} (see Equation~\ref{eqn:ritchey}) and the direct determination of the CMB temperature \citep{Fixsen2009}. This possibly suggests that individual model fits might be over-fitting to the noise, while a single joint fit is less affected by the parameter minimisation; by performing a joint fit, we can ensure that any one spectrum is not causing the data to be over-fit, thereby biasing the $\ronezero$ value. Our estimated $T_{12}$ values are significantly elevated relative to the CMB value, indicating that either local sources are contributing to the relative level populations, or the (much weaker) $R(2)$ and $P(2)$ lines are being fit to the noise. Finally, we note that the column density ratio of $^{13}{\rm C}^{14}{\rm N}$ is $\ronezero=0.467\pm0.077$, corresponding to an excitation temperature $T_{01}(^{13}{\rm CN}) = 2.80 \pm0.25~{\rm K}$, which is consistent with the $^{12}{\rm CN}$ results.

\subsection{Future improvements}

Improvements to the future generations of ultra-high resolution spectrographs will hopefully allow us to accurately model the LSF, and reduce the intrinsic (systematic) scatter that is currently present in the CN excitation measurements. In this section, we briefly reflect on the key properties that are needed to significantly improve upon these measurements in the future.

Similar to previous studies, we find that the LSF plays an important role in the determination of the excitation temperature of unresolved CN lines \citep{Palazzi1992,Roth1993,Roth1995,Slyk2008,Ritchey2011}. To improve the measurement reliability, we suggest that the absorption lines need to be fully resolved. This would allow the cloud model parameters to be measured with greater accuracy, including the possibility of identifying multiple blended components along some lines of sight. Assuming that unresolved velocity components do not affect the systems in our sample, we find that the typical Doppler parameter of CN absorption lines is $\langle b \rangle = 0.61 \pm 0.14~{\rm km~s}^{-1}$ (corresponding to a line ${\rm FWHM}\simeq1.0~{\rm km~s}^{-1}$). The lowest Doppler parameter that we measure is towards one component of HD\,149757 ($\zeta$~Oph), $b=0.36\pm0.03$ (equivalent to ${\rm FWHM}=0.58~{\rm km~s}^{-1}$). The highest value measured is towards the strongest CN absorber (HD\,73882), with $b=0.84\pm0.02$, although we note that this value may be biased due to line saturation combined with the LSF uncertainty. Given that at least some CN Doppler parameters in the Milky Way are lower than ${\rm FWHM}\approx0.6~{\rm km~s}^{-1}$, we recommend that a spectral resolution $R>500\,000$ is required to obtain reliable measurements of the CN excitation temperature. As an aside, we also note that the typical Doppler parameter that we derive in this work differs by $\sim40$ percent from the value that was uniformly assumed by \citet[][$b=1~{\rm km~s}^{-1}$]{Slyk2008}. This difference likely explains the excitation excess reported by these authors (see also, \citealt{Ritchey2011}).

We have also found that the LSF does not significantly impact the results when the column density of the lower level is $N_{l}\lesssim10^{12.5}~{\rm cm^{-2}}$. This is perhaps the ideal column density regime to use for reliable $T_{01}$ determinations, when analysing the $^{12}{\rm C}^{14}{\rm N}$ lines. We also point out that $^{12}{\rm C}^{14}{\rm N}$ absorption line systems in the regime $N_{l}\gtrsim10^{13}~{\rm cm^{-2}}$ are well-suited to measure $T_{01}$ from the $^{13}{\rm C}^{14}{\rm N}$ absorption lines. The benefit of using the heavier isotope is that the energy level difference is $\sim 5$ percent closer to the peak of the CMB than $^{12}{\rm C}^{14}{\rm N}$, and the level populations are therefore even more strongly determined by the temperature of the CMB photons. The strongest $^{12}{\rm C}^{14}{\rm N}$ lines in principle allow for a high precision determination of $T_{01}$, but this precision must be matched with an accurate determination of the cloud model. In particular, the Doppler parameter must be well-measured; this is possible by using a combination of both weak and strong absorption lines arising from the same lower level, and fitting the lines of different CN bands simultaneously.

In addition to acquiring new data of higher spectral resolution, we suggest that the S/N of the collected data should allow for the $R(1)$ and $P(1)$ absorption lines to be detected with high confidence (i.e. the column density should be measured with S/N=100). There are just two sightlines that we have analysed with ${\rm S/N} < 100/{\rm pixel}$. One of these sightlines (HD\,152236) exhibits an excitation temperature that is (unphysically) below the CMB temperature by $\sim1.4\sigma$. The sightline with the lowest S/N of our sample (HD\,152270) has an excitation temperature that is significantly above ($\sim2\sigma$) the CMB temperature. It is important to assess such elevated cases carefully, since an excitation temperature that exceeds the CMB value could be explained by local sources of excitation. S/N therefore plays an important role in obtaining reliable measurements.

In principle, the amount of local excitation can be determined directly from millimeter observations of CN rotational line emission \citep{Penzias1972}, provided that the source uniformly fills the antenna beam. There are two sightlines in our sample with millimeter observations available. HD\,149757 ($\zeta$~Oph) has a $2\sigma$ upper limit, $\Delta T_{\rm loc}<0.062~{\rm K}$ \citep{Crane1989}. We note that our data of HD\,149757 are among the highest spectral resolution and highest S/N available, and yet the excitation temperature ($T_{01}=2.804\pm0.032~{\rm K}$) is marginally discrepant ($1.8\sigma$) with the CMB value, given the aforementioned upper limit on the level of local excitation. On the other hand, HD\,27778 has an estimated local excitation contribution $\Delta T_{\rm loc}=0.02\pm0.02~{\rm K}$ \citep{Roth1993}. However, the excitation temperature that we report herein for this sightline ($T_{01}=2.726\pm0.022~{\rm K}$) is in excellent agreement with the CMB temperature, and does not require local sources to be brought into agreement with the CMB value. As the community pushes towards higher precision, it will be necessary to obtain high quality data in the optical and millimeter bands to firmly assess the relative importance of local excitation and systematic uncertainty.

\section{Conclusions}
\label{sec:conc}

We have analysed ultra-high resolution spectra of CN \band\ (0,0) vibronic band absorption lines to determine the typical excitation temperature of diffuse molecular clouds in the Milky Way. In the Milky Way environment, the first rotationally excited state of CN is largely dominated by the absorption of CMB photons, and the excitation temperature is expected to be very close to the CMB temperature. The main goals of this paper include:
(1) understand the systematic limitations of using CN to infer the CMB temperature;
(2) estimate (or place a limit on) the contribution of local sources to the rotational excitation of CN; and
(3) if local sources are subdominant, report a robust measure of the CMB temperature.
Following a careful analysis of eight Milky Way sightlines, we draw the following conclusions:

\smallskip

(i) We report a $3\%$ determination of the excitation temperature of CN in diffuse molecular clouds, $T_{01}=2.769^{+0.084}_{-0.072}~{\rm K}$, based on a sample size of eight CN absorbers. Our determination is consistent with the direct determination of the CMB temperature reported by \citet{Fixsen2009}, supporting previous works that also conclude the CN level populations are dominated by excitation of CMB photons. Local sources of excitation appear to contribute very little to the level populations.

\smallskip

(ii) We investigate several possible causes of systematic uncertainty, including the LSF, initialisation bias, fine-structure of the absorption lines, the zero-level and continuum level of the data, and blended components. We also perform a series of self-consistency checks, such as fitting the level populations of separate spectra and separate absorption components. We have also conducted our analysis blind, to reduce the impact of human bias on the results. Despite the care that was taken in the analysis, we find an excess dispersion in the reported excitation temperature measurements that can be explained by an unaccounted for systematic. We estimate that the column density ratio $\ronezero$ is uncertain at the $\sim10$ percent level, given the current data.

\smallskip

(iii) We have also performed a simultaneous joint fit to all absorbers, requiring that all CN absorption components have an identical excitation temperature. The simultaneous analysis of all absorbers yields a typical CN excitation temperature $T_{01}=2.725\pm0.015~{\rm K}$, which is consistent with the CMB temperature, but with five times higher precision than analysing all absorbers separately. We suggest that a simultaneous joint fit might alleviate any one system from being overfit, and we outline a future observing strategy to test this possibility.

\smallskip

(iv) We find that typical CN absorption clouds have total Doppler parameters of $\langle b \rangle=0.61\pm0.14~{\rm km~s}^{-1}$. In order to make further progress on the determination of the typical CN excitation temperature of Milky Way diffuse molecular clouds, we suggest that future observations should be acquired with high S/N data (${\rm S/N}\gtrsim200/{\rm pixel}$, such that the $R(1)$ and $P(1)$ lines are detected at S/N=100) and attempt to fully resolve the absorbing clouds. Such a demand would require an exceptionally high resolution spectrograph ($R\gtrsim10^{6}$).

\smallskip

(v) As an added bonus of this work, we detected isotopic CN absorption towards HD\,62542, including a marginal detection of $^{12}{\rm C}^{15}{\rm N}$. The isotopic abundances that we measure along this sightline are $^{12}{\rm C}{\rm N}/^{13}{\rm C}{\rm N} = 55.4 \pm 8.5$ and ${\rm C}^{14}{\rm N}/{\rm C}^{15}{\rm N} = 790 \pm 430$.

\smallskip

The present day CMB temperature is currently the most precisely measured cosmological quantity. Furthermore, its redshift evolution has been measured using a variety of techniques covering most of the expansion history of the Universe (out to $z\simeq6.3$), and the agreement with the standard cosmological model is exceptional. Given that future facilities aim to measure the CMB temperature using CN at higher redshift, it is important that we first understand the systematic limitations of this approach using exquisite data that we can collect from our Galactic neighbourhood. Such a study will likely require a purpose-built optical spectrograph with a resolution that is higher than has ever been used for astronomical observations.

\section*{Acknowledgements}

We would like to thank the referee, Adam Ritchey, for a thorough and constructive report that improved the clarity of our rotational excitation discussion, the line profile fitting, and for offering a comparison between our independent works. We warmly thank Ian Crawford, for sharing the D-8 data tapes of their UHRF observations, and for their advice and recollections about the observations. We also thank John Lucey for recollections about the \textsc{vmsbackup} command that was used to store data on Exabyte tapes some three decades ago!
During this work, RJC was funded by a Royal Society University Research Fellowship. RJC acknowledges support from STFC (ST/T000244/1, ST/X001075/1). 
This work has been supported by Fondazione Cariplo, grant No 2018-2329. This work used the DiRAC@Durham facility managed by the Institute for Computational Cosmology on behalf of the STFC DiRAC HPC Facility (www.dirac.ac.uk). The equipment was funded by BEIS capital funding via STFC capital grants ST/K00042X/1, ST/P002293/1, ST/R002371/1 and ST/S002502/1, Durham University and STFC operations grant ST/R000832/1. DiRAC is part of the National e-Infrastructure.
This research has made use of NASA's Astrophysics Data System.
Based on observations made with ESO Telescopes at the La Silla Paranal Observatory under programme ID : 0102.C-0699(A).
For the purpose of open access, the author has applied a Creative Commons Attribution (CC BY) licence to any Author Accepted Manuscript version arising from this submission.

\section*{Data Availability}

All of the data analysed in this paper are publicly available in the astronomy data archives that are linked in this paper. Reduced spectra are available from the lead author upon request.



\bibliographystyle{mnras}
\bibliography{tcmb} 








\bsp	
\label{lastpage}
\end{document}